%% file: main.tex
\documentclass[10pt,aps,prx,twocolumn]{revtex4-2}
\usepackage[linkcolor = blue, citecolor = blue, urlcolor = blue, colorlinks = true]{hyperref}
\usepackage[usenames,dvipsnames]{xcolor}
\usepackage[normalem]{ulem}
\usepackage{graphicx}
\usepackage{listings}
\usepackage{stmaryrd}
\usepackage{mathrsfs}
\usepackage{amssymb}
\usepackage{amsmath}
\usepackage{gensymb}
\usepackage{float}
\usepackage{bbold}
\usepackage{bm}

\newcommand{\R}{\mathfrak{R}}

\hypersetup{
    colorlinks=true,
    linkcolor=blue,
    citecolor=red}

\begin{document}

\title{Stretching theory of Hookean metashells}

\author{Luca Giomi}
\email{giomi@lorentz.leidenuniv.nl}
\affiliation{Instituut-Lorentz, Universiteit Leiden, P.O. Box 9506, 2300 RA Leiden, The Netherlands}

\date{\today}

\begin{abstract}
Despite being governed by the familiar laws of Hookean mechanics, elastic shells patterned with an internal structure (i.e. metashells) exhibit a wealth of unusual mechanical properties with no counterparts in unstructured materials. Here I show that much of this behavior can be captured by a real-valued analog of the inhomogeneous Schr\"odinger equation, with the lateral pressure experienced by the internal structure in the role of the wave function. In the fine structure limit -- i.e. when the length scale associated with the internal structure is much smaller than the local radius of curvature -- this approach reveals the existence of localized states, in which elastic deformations are prevented to diffuse away from their origin, thereby allowing the internal structure to smoothly adapt to the intrinsic geometry of the metashell. Leveraging on an analogy with scattering states in nearly free electrons, I further show that periodic metashells, obtained from the repetition of the same structural unit periodically in space, support elastic Bloch waves, corresponding to stationary periodic configurations of the internal structure and characterized by a geometry-dependent band structure. When applied to crystalline monolayers, this approach provides a generalization of the elastic theory of interacting topological defect to compressible systems.
\end{abstract}

\maketitle

Additive manufacturing -- i.e. the fabrication of physical objects from three-dimensional computer models via layer-by-layer deposition -- has experienced an enormous advancement in the past two decades and has now entered in the realm of commonly available technologies: from research-oriented applications to large-scale industrial productions~\cite{Lipson:2013}. Among the wealth of opportunities opened by this ongoing paradigm shift, {\em metamaterials} possibly represent one of the most promising and interesting examples, both in terms of conceptual depth and technological impact. Metamaterials are engineered materials whose small-scale structure is specifically designed to obtain optical, electromagnetic or mechanical responses that are uncommon in nature. In mechanics, in particular, this effort led to the discovery a variety of fascinating phenomena, where the local geometry and kinematics of the system's building blocks cooperate to give rise to structural and acoustic properties hardly found in naturally occurring materials. Examples include the existence of protected edge modes in mechanical topological insulators \cite{Kane:2014,Chen:2014,Huber:2016}, the possibility of programming shape via elastic instabilities~\cite{Shim:2012,Florijn:2014} or small-scale patterning~\cite{Choi:2019}, as well as various realizations of negative elastic moduli~\cite{Lakes:1987,Lakes:1987,Bertoldi:2010,Nicolaou:2012,Coulais:2015}. 

\begin{figure}[b]
\centering
\includegraphics[width=\columnwidth]{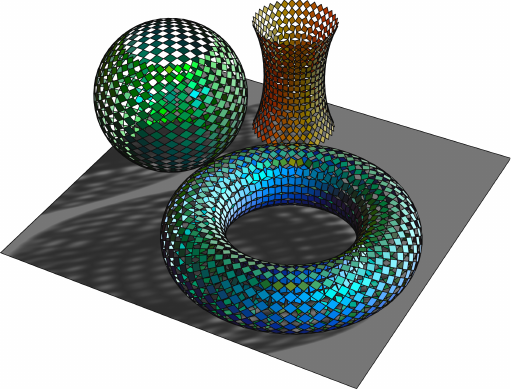}
\caption{\label{fig:illustration}Examples of metashells obtained by embedding a deployable kirigami lattice on three surfaces of revolution: i.e. a torus (blue), a spherical barrel (green) and a pseudospherical barrel (green). The latter two surfaces feature constant positive and negative Gaussian curvature respectively, while Gaussian curvature of the torus varies from positive (outside) to negative (inside).}	
\end{figure}

Because of their intrinsically discrete nature, which normally consists of the repetition of the same structural motif periodically in space, much theoretical research on mechanical metamaterials has so far focused on the properties of individual building blocks and how these can be integrated in order to achieve specific functionalities at the macroscopic scale~\cite{Coulais:2016,Rafsanjani:2017}. By contrast, limited work has been done towards developing continuum theories of mechanical metamaterials, where the local configuration of the system's internal structure and the resulting stress distribution can be described in terms of one or more smoothly varying fields~\cite{Bertoldi:2007,Stoop:2015}. This article aims at progressing the latter approach, by leveraging on a few classic and yet powerful concepts of differential and integral geometry, whose connection -- albeit not obvious {\em a priori} -- naturally emerges within the solid framework of Hookean elasticity. 

To this end, we focus on a specific class of elastic shells endowed of an arbitrary internal structure, hereafter referred to as {\em microstructure}. These shells are assumed to have a fixed but non-trivial shape, to which the microstructure adapts by deforming along the pathway of in-plane transformations built into its specific architecture. An simple realization of this setting is illustrated in Fig.~\ref{fig:illustration} and consists of a common type of {\em kirigami} lattice, whose vertices are forced to lie on a smooth surface. For sake of conciseness, we will refer to these metamaterial shells as {\em metashells}. In the following, I will demonstrate that both the local configuration of the microstructure and the lateral pressure this experiences can be obtained from the solution of a single scalar partial differential equation, reflecting the frustration arising from the incompatibility between the microstructure's kinematic and the geometry of the shell. 

\noindent {\\[5pt]\bf Hooekean metashells}\\
Let us then consider a metashell featuring a generic microstructure and whose mid-surface's metric is given by ${\rm d}s^{2}=g_{ij}{\rm d}x^{i}{\rm d}x^{j}$, with $g_{ij}$ the metric tensor and $\{x^{1},x^{2}\}$ generic contravariant coordinates, when undeformed. Let us further assume that the microstructure obeys to the standard Hooke law. That is
\begin{equation}\label{eq:hooke}
\sigma_{ij} = \lambda u_{k}^{k}g_{ij}+2\mu u_{ij}\;,
\end{equation}
where $\sigma_{ij}$ and $u_{ij}$ are the stress and strain tensors respectively, while $\lambda$ and $\mu$ are the two Lam\'e coefficients and, together with the other two-dimensional elastic moduli, are tabulated in Table~\ref{tab:elastic_moduli} for convenience. Next, let us consider a generic deformation $\bm{r}\to\bm{r}'=\bm{r}+\delta\bm{r}$, whose effect is to change the local Gaussian curvature: i.e. $K \to K'=K+\delta K$. After lengthy algebraic manipulations~\cite{SI}, it is possible derive the following equation for the lateral pressure at the linear order in $\delta\bm{r}$: 
\begin{equation}\label{eq:pressure}
\left(\frac{2}{Y}\,\nabla^{2}+\frac{K}{B}\right)P = \delta K\;,
\end{equation}
with $\nabla^{2}=\nabla^{i}\nabla_{i}$ is the Laplace-Beltrami operator on the surface,$Y$ the Young modulus and $B$ the bulk modulus (see Table~\ref{tab:elastic_moduli}). Several comments are in order.
\begin{table}[t!]
\begin{ruledtabular}
\begin{tabular}{cccc}
{\rm Modulus} & $(\lambda,\mu)$                          & $(\nu,Y)$                 & $(B,G)$             	\\[5pt]
\hline
$\lambda$     & $\lambda$                                & $\frac{\nu Y}{1-\nu^{2}}$ & $B-G$  				\\[5pt]
$\mu$ 	      & $\mu$                                    & $\frac{Y}{2(1+\nu)}$      & $G$    				\\[5pt]
$\nu$         & $\frac{\lambda}{\lambda+2\mu}$           & $\nu$                     & $\frac{B-G}{B+G}$ 	\\[5pt]
$Y$ 	      & $\frac{4\mu(\lambda+\mu)}{\lambda+2\mu}$ & $Y$                       & $\frac{4BG}{B+G}$  	\\[5pt]
$B$           & $\lambda+\mu$                            & $\frac{Y}{2(1-\nu)}$      & $B$                 	\\[5pt]
$G$           & $\mu$                                    & $\frac{Y}{1+\nu}$         & $G$                 	\\
\end{tabular}
\end{ruledtabular}
\caption{\label{tab:elastic_moduli}Two-dimensional elastic moduli. The quantities $\lambda$ and $\mu$ are the first and second Lam\'e coefficients, $\nu$ and $Y$ Poisson ratio and Young and $B$ and $G$ the bulk and shear modulus respectively.} 
\end{table}

First, Eq.~\eqref{eq:pressure} is exact within the limit of validity of linear Hookean elasticity, but, to the best of the writer's knowledge, was not known in this generic form. Conversely, in a number of special cases, Eq.~\eqref{eq:pressure} can be casted in forms which are routinely used in elastic theories of plates and shells. For instance, if the mid-surface is flat when undeformed -- i.e. $g_{ij}=\delta_{ij}$ and $K=0$ -- the lateral pressure can be expressed in terms of the so-called Airy stress function and Eq.~\eqref{eq:pressure} reduces to the stress equation of the F\"oppl-von K\'arm\'an theory \cite{Audoly:2010}, as well as its generalization to non-Euclidean plates \cite{Klein:2007,Efrati:2009,Liang:2009}, thin sheets and frames~\cite{Paulsen:2016,Moshe:2019} and other type of structured surfaces~\cite{Seung:1988,Bar:2020}. In all these examples, however, the undeformed configuration of the mid-surface is assumed flat and the second term on the left-hand side vanishes, while the source term on the right-hand side originates from non-linear out-of-plane contributions to the strain tensor rather than from the linear contributions arising from the shell's pre-existing curvature. If the mid-surface undeformed configuration consists instead of a sphere of radius $R$ and Gaussian curvature $K=1/R^{2}$, Eq.~\eqref{eq:pressure} yields the stress equation of spherical shells (see Sec. S1 in Ref.~\cite{SI} and Ref.~\cite{Niordson:1985}). Notice that, because of the non-commutativity of covariant derivatives on intrinsically curved surfaces, it is impossible to exactly parametrize the stress tensor in terms of a single Airy stress function unless $K$ is constant.

Second, Eq.~\eqref{eq:pressure} is formally identical to the screened Poisson equation~\cite{Chen:2016}, with the radius of curvature of the undeformed shell playing the role of Debye's length. Yet, unlike electrostatic screening, which always results in a damping of Coulomb's interactions, the peculiar form of {\em elastic screening} entailed in Eq.~\eqref{eq:pressure}, can lead to both a damping and an enhancement of the elastic interactions, depending whether the sign of $K/B$ is negative or positive. We will come back to this concept later with examples of kirigami metashells. 

Third, the aforementioned screening effect crucially depends on the shell compressibility, embodied in the bulk modulus $B$. In incompressible shells, where $B\to\infty$, the second term on the left-hand side of Eq.~\eqref{eq:pressure} vanishes and elastic interactions are no longer damped or enhanced. This effect is, therefore, expected to be minor in the majority of natural occurring materials as well as in the most common construction materials -- such as steel and concrete -- being these close to incompressible. By contrast, the majority of mechanical metamaterials, are specifically designed to be highly compressible, for instance by means of reentrant unit cells~\cite{Bertoldi:2010,Coulais:2015}. Elastic screening is, therefore, expected to be significant in this case. 

To make progress, Eq.~\eqref{eq:pressure} must be closed by expressing $\delta K$ in terms of the lateral pressure $P$. This can be achieved by interpreting $K'$ as an {\em effective} Gaussian curvature arising from the deformation of the microstructure. A possible route towards this mapping can be found in a seminal result of integral geometry, known as Hadwiger’s characterization theorem~\cite{Hadwiger:1957}. This asserts that any motion-invariant {\em valuation} -- i.e. a scalar integral quantity reflecting the shape of a geometric object -- of a $d-$dimensional convex body can be expressed as the linear combination of $d+1$ irreducible valuations known as Minkowski functionals~\cite{Mecke:1998}. When $d=2$, the corresponding Minkowski functionals are proportional to the body's area ($\mathcal{A}$), perimeter ($\mathcal{L}$) and Euler characteristic ($\chi$). The fundamental assumption behind our theory is that the Gaussian curvature $K'$ of a local {\em patch}, here defined a portion of metashell where the lateral pressure is approximatively uniform, is effectively a valuation reflecting the shape of the deformed patch. From this it follows that
\begin{equation}\label{eq:hadwiger}
K' = \alpha_{0}\mathcal{A}'+\alpha_{1}\mathcal{L}'+\alpha_{2}\chi'\;,
\end{equation}
where the prime denotes, as before, deformed configurations. The coefficients $\alpha_{0}$, $\alpha_{1}$ and $\alpha_{2}$ depend on the specific architecture and kinematics of the microstructure, but are not independent, since elastic deformations preserve the Euler's characteristic (i.e. $\chi'=\chi$). Furthermore, $K'=0$ when $\mathcal{A}'=\mathcal{A}$ and $\mathcal{L}'=\mathcal{L}$ by construction. This allows one to express $K'$ as a sole function of the area and perimeter variations: i.e. $\delta\mathcal{A}=\mathcal{A}'-\mathcal{A}$ and $\delta\mathcal{L}=\mathcal{L}'-\mathcal{L}$. These variations, on the other hand, depends upon the later pressure experienced by the microstructure, so that $\delta\mathcal{A}=-\mathcal{A}\Pi$ and $\delta\mathcal{L}=-(\Theta/\mathcal{L})\mathcal{A}\Pi$~\cite{Giomi:2013}, with $\Pi=P/B$ a dimensionless form of the lateral pressure and $\Theta$ the global turning angle, that is the angle the tangent vector rotates in one loop along the boundary (e.g. $\Theta=2\pi$ for simple closed curves). Thus, Eq.~\eqref{eq:hadwiger} can be cast in the simple form
\begin{equation}\label{eq:keff}
K' = -K_{0}\Pi\;,
\end{equation}
where $K_{0}=(\alpha_{0}+\alpha_{1} \Theta/\mathcal{L})\mathcal{A}$ a constant depending on the structural and kinematic properties of the microstructure (see Sec. S2 in Ref.~\cite{SI} for the complete derivation). This constant can be treated as a material parameter and inferred from experiments or explicitly computed upon assuming a specific deformation model of the shell microstructure. In Sec. S3 of Ref.~\cite{SI}, I elucidate this concept by modeling a deformed metashell patch a positionally-constrained $n-$sided geodesic polygon, for which $\alpha_{0}=2n^{2}/(\mathcal{A}\mathcal{L}^{2})$ and $\alpha_{1}=12n^{2}/\mathcal{L}^{3}$. Finally, replacing Eq.~\eqref{eq:keff} in Eq.~\eqref{eq:pressure} one finds
\begin{equation}\label{eq:shape}
\beta\nabla^{2}\Pi+(K+K_{0})\Pi + K = 0\;,
\end{equation}
where $\beta=2B/Y$ is a dimensionless compressibility factor. Eq.~\eqref{eq:shape} is a real-valued analog of the time-independent inhomogeneous Schr\"odinger equation and represents the central equation of our theory. Together with knowledge of the unit cells' architecture and of the shell's geometry and boundary pressure, Eq.~\eqref{eq:shape} allows one to calculate the configuration of the lateral pressure and of the microstructure over the entire shell.

\begin{figure*}[t!]
\centering
\includegraphics[width=\textwidth]{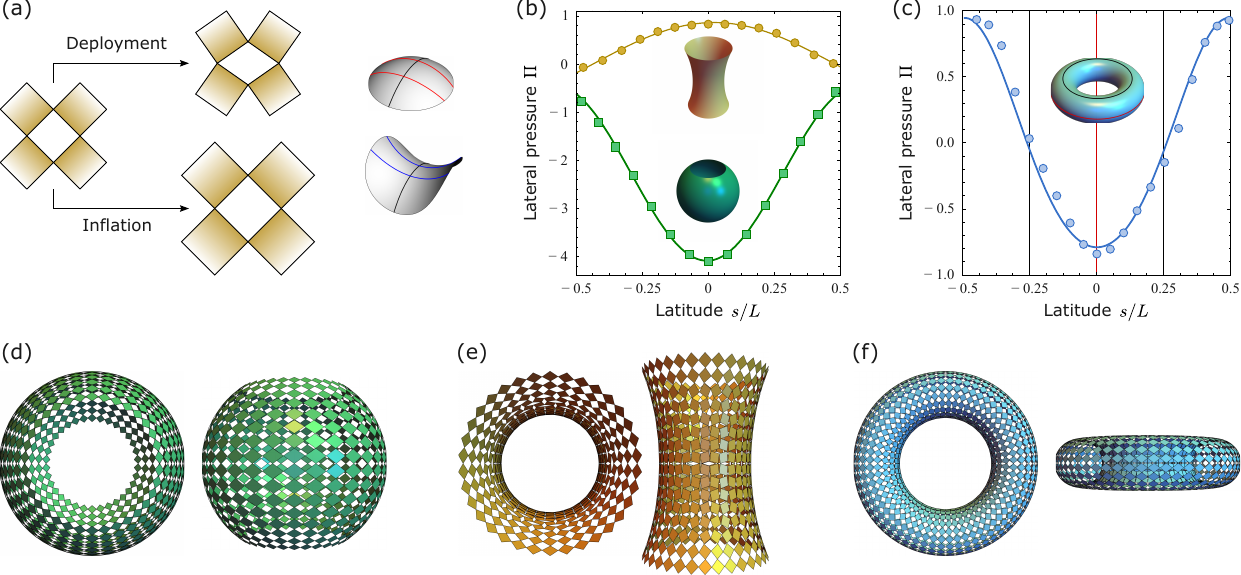}
\caption{\label{fig:data}Stretching of kirigami methashells. (a) On the left, an illustration of the fundamental deformation modes of a unit cell: i.e. deployment and inflation. On the right, an example of how the underlying Gaussian curvature of a surface causes the geodesic to converge (top, $K>0$) or diverge (bottom, $K<0$). (b,c) Spatial configuration of the dimensionless later pressure $\Pi$ obtained from numerical simulations of a spring model of kirigami metashell, with spherical/pseudospherical and (b) toroidal geometry (see Sec.~S5 of Ref.~\cite{SI} for details). The red and black lines in in panel (c) mark the positions of the {\em external} equator and the circles, located at $s= \pm L/4$, where the Gaussian curvature changes in sign. (d-f) Top and side views of the simulated metashells.}	
\end{figure*}

Before looking at specific examples, it is important to notice that Eq.~\eqref{eq:shape} admits a general asymptotic solution in the limit of an infinitely fine microstructure. This is because, regardless of its specific form, $K_{0} \sim 1/\mathcal{L}^{2}$, thus diverges for $\mathcal{L}\to 0$, that is when the size of a patch -- proportional to that of the individual unit cells comprising the shell's microstructure -- is much smaller than the local radius of curvature. In this limit, the first two terms on the left-hand side of Eq.~\eqref{eq:shape} become sub-leading and an asymptotically exact general solution can be constructed in the form
\begin{equation}\label{eq:k0_to_infty}
\Pi \approx -\frac{K}{K_{0}}\;,
\end{equation}
from which $K' \to K$ and $\Pi \to 0$ in the bulk. Thus, the microstructure adapts progressively more efficiently to the shell's geometry as its resolution becomes higher and higher, while the lateral pressure decreases and vanishes in the infinite resolution limit. In the following we will see how this result, albeit intuitive, originates from a  peculiar phenomenon of elastic localization, with no counterparts in unstructured materials. 

\noindent {\\[5pt]\bf Finite-resolution metashells}\\
To make progress, we look for explicit solutions of Eq.~\eqref{eq:shape} for the three metashells illustrated in Fig.~\ref{fig:illustration} and featuring a planar deployable kirigami pattern, consisting of a checkerboard lattice of solid and empty squares held together by hinges (Fig.~\ref{fig:data}a), embedded on three surfaces of revolutions: i.e. a spherical and a pseudospherical barrel, having constant positive and negative Gaussian curvature respectively, and a torus, whose Gaussian curvature varies from positive (outside) to negative (inside). By virtue of the substrates' azimuthal symmetry, in all three cases we can assume $\Pi$ to depend solely on the latitudinal coordinate, which, in turn, can be conveniently expressed in terms of geodesic distance $s$ from the surfaces' equator (external equator for toroidal metashells), so that $-L/2\le s \le L/2$, with $L$ the length of the metashell along any of its meridians. A general solution of Eq.~\eqref{eq:shape} can then be expressed in the form $\Pi=\Pi_{h}+\Pi_{p}$, where $\Pi_{h}=\Pi_{h}(s)$ is the solution of the homogeneous equation associated with Eq.~\eqref{eq:shape} and $\Pi_{p}=\Pi_{p}(s)$ a particular solution. As detailed in Sec. S5 of Ref.~\cite{SI}, the latter can be expressed as
\begin{gather}\label{eq:general_solution}
\Pi_{p} = \sum_{n=0}^{\infty}(-1)^{n+1}\mathscr{D}^{2n}\left(\frac{K}{K_{0}+K}\right)\;,	
\end{gather}
under the assumption that $\beta<K_{0}L^{2}$. Regardless the specific magnitude of the dimensionless compressibility $\beta$, this inequality is always satisfied for a sufficiently fine microstructure, hence in any realization of metashells which can be reasonably addressed by means of a continuum approach. In Eq.~\eqref{eq:general_solution}, $\mathscr{D}^{2n}(\cdots)$ denotes $n$ applications of the differential operator $\mathscr{D}^{2}(\cdots) = \beta/(K_{0}+K)\,\nabla^{2}(\cdots)$.

Now, in the case of spherical and peseudospherical barrels, where $K=\pm 1/R^{2}$ with $R$ a constant radius of curvature, the only non-vanishing term of the summation on the right-hand side of Eq.~\eqref{eq:general_solution} corresponds to $n=0$, from which $\Pi_{p}=-1/(1\pm K_{0}R^{2})$. The solution $\Pi_{h}$ of the associated homogeneous equation with boundary conditions $\Pi(-L/2)=\Pi(L/2)=\Pi_{\rm ext}$, with $\Pi_{\rm ext}$ a constant externally applied pressure, can instead be expressed in terms of Legendre functions (see Sec. S5 of Ref.~\cite{SI}). As the fine structure limit is approached and $K_{0}R^{2}\gg 1$, the latter reduces to $\Pi_{h} \approx (\Pi_{{\rm ext}}-\Pi_{p})\cos (k_{\pm}s)/\cos(k_{\pm}L/2)$, with $k_{\pm}^{2}=(K_{0}\pm 1/R^{2})/\beta$. This is shown in Figs.~\ref{fig:data}b and \ref{fig:data}c, together with data points obtained from a numerical minimization of a spring model of the same metashells (see Figs.~\ref{fig:data}d-f and Sec. S6 of Ref.~\cite{SI} for details). For spherical and pseudospherical barrels, these plots provide an explicit example of the anticipated curvature-mediated screening mechanism. In both spherical (green tones) and pseudospherical (yellow tones) barrels, the enlargement (reduction) of the metashells' width introduces geometrical frustration, by preventing the microstructure from keeping its original conformation away from the boundary. Yet, while in pseudospherical barrels this is compensated by a {\em deployment} of the kirigami microstructure, the latter does not feature deformation pathways that could accommodate the positive Gaussian curvature of spherical barrels without significant stretching, thus resulting in a progressive {\em inflation} of the unit cells away from the boundaries (see Fig.~\ref{fig:data}a). The same organization is found in toroidal metashells (Figs.~\ref{fig:data}c and \ref{fig:data}f), where the Gaussian curvature varies from positive to negative along meridians. In this case, however, $\Pi_{h}=0$ in the absence of pre-stresses and the total lateral pressure vanishes: i.e. $\int {\rm d}A\,\Pi=0$. Notice that, while the Gaussian curvature is larger in magnitude in the interior of the torus, the lower energy cost of deployment compared to inflation causes the lateral pressure to be approximatively the same on both equators, i.e: $\Pi(0) \approx -\Pi(L_/2)$. 

\noindent {\\[5pt]\bf Elastic localization\\}
The phenomenology reported so far in the context metashells with finite resolution allows us to ascribe the fine structure limit, Eq.~\eqref{eq:k0_to_infty}, to a phenomenon of elastic localization. On a surface, the sign of the Gaussian curvature determines the dispersion of the geodesic flow, in a way non dissimilar to how the refractive index of an optical medium determines how light is bent while traveling through it~\cite{Santangelo:2007}. To illustrate this concept, let $\ell=\ell(s)$ be the distance between two geodesics (red and blue lines in Fig.~\ref{fig:data}a), parametrized in terms of the arc-length $s$ along either geodesic, and perpendicular to a third geodesic (black line in Fig.~\ref{fig:data}a) at $s=0$. Close to the origin, $\ell$ obeys Jacobi's equation: i.e. $(\partial_{s}^{2}+K)\ell=0$~\cite{DoCarmo:2016}. Thus, while moving away from the intersections, the geodesics diverge for $K<0$ and converge for $K>0$, thereby inducing a local compression or extension of the shell (see Figs.~\ref{fig:data}d-f). In addition, the same intrinsic geometry mediates the propagation of deformations originating at a distance, thus leading to the screening mechanism embodied in the second term on the left-hand side of Eq.~\eqref{eq:pressure}. Specifically, for $K<0$, the divergence of the geodesic flow causes a dispersion of the strain field lines, hence a damping of the lateral pressure, while their focusing, for $K>0$, drives a pressure enhancement. The latter does not depend on the presence of the microstructure and is common to all Hookean shells alike, with the local radius of curvature serving as a screening length: i.e. $\xi= \pm \sqrt{\beta/|K|}$. This picture is dramatically affected by the kinematic of the microstructure, which, by virtue of Eq.~\eqref{eq:shape}, determines a renormalization of the screening length: i.e. $\xi=\sqrt{\beta/|K_{0}+K|}$. In the fine structure limit, when $K_{0}\to \infty$, $\xi \to 0$ irrespectively of the substrate's intrinsic geometry and an elastic deformation caused by the curvature remains {\em localized} at its origin, where it is entirely compensated by a local rearrangement of the microstructure. In this respect, the general asymptotic solution given by Eq.~\eqref{eq:k0_to_infty} provides an example of a fully localized elastic state, where deformations are prevented to diffuse away from their origin.

\begin{figure}[t]
\centering
\includegraphics[width=\columnwidth]{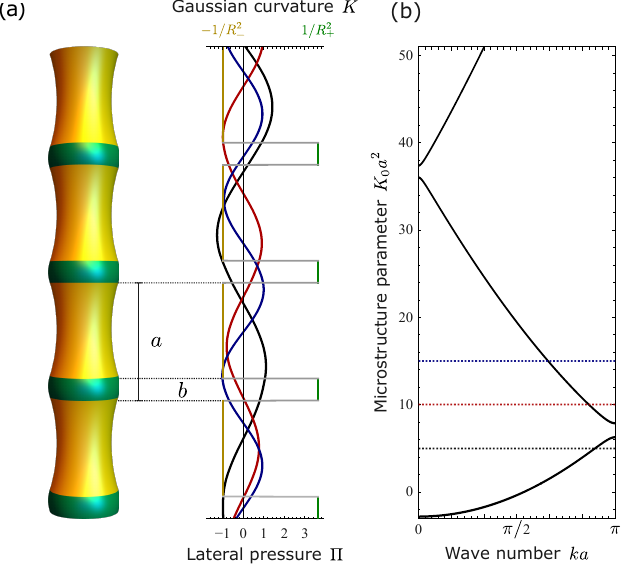}
\caption{\label{fig:metatubes}Bloch waves in periodic metatubes. (a) Example of periodic tubular surface obtained upon connecting spherical and pseudospherical barrels so to guarantee the continuity of the tangent plane, while the Gaussian curvature is piecewise constant function of the geodesic latitude switching in between $-1/R_{-}^{2}$ and $1/R_{+}^{2}$. On the right, a portion of a Bloch wave solution of Eq.~\eqref{eq:shape} for $\beta=1$, $R_{-}=a$, $R_{+}/R_{-}=0.52414$ and $b/a=0.18708$ and three different $K_{0}$ values. (b) Band structure associated with the microstructure of Bloch waves for the same parameter values of panel (a). The horizontal dotted lines mark the $K_{0}$ values of the three solutions.}	
\end{figure}

\noindent {\\[5pt]\bf Bloch waves in metatubes}\\
Our discussion so far has made little use of the analogy between Eq.~\eqref{eq:shape} and the inhomogeneous Schr\"odinger equation. The main limitation preventing such a formal analogy from becoming substantial lies on the fact that, unlike the later pressure $\Pi$, the absolute magnitude of the wave function $\psi$ is constrained by the probabilistic interpretation of quantum mechanics. That is $\int_{\Omega} {\rm d}V\,|\psi|^{2}=1$, in any finite volume domain $\Omega$. This, for instance, excludes from the elastic boundary value problem the existence of {\em bound states} with discrete energy levels. The same limitation, however, does not apply to {\em scattering states}, characterized by a continuous energy spectrum and a non-normalized wave function. To provide an example of the latter, here we discuss the case of an infinite periodic {\em metatube}, consisting of the repetition of the same structural unit, comprising a pseudospherical barrel, of curvature $K=-1/R_{-}^{2}$ and length $L=b$, and a spherical barrel, of curvature $K=1/R_{+}^{2}$ and length $L=a-b$, connected to each other so to guarantee the continuity of the tangent plane (see Fig.~\ref{fig:metatubes}a). In the fine structure limit, when $K_{0}R_{\pm}^{2}\gg 1$, the force balance condition expressed by Eq.~\eqref{eq:shape} reduces then to a inhomogeneous variant of Kronig's and Penney's model of electrons in one-dimensional crystals~\cite{Kronig:1931}, whose solution can be comprehensively analyzed in the framework of Floquet theory~\cite{Yakubovich:1975,Slane:2011,Kollmitzer:2011}. As in the analogous quantum mechanical problem, the existence of bounded solutions here depends on the structure of the solution of the associated homogeneous equation~\cite{Slane:2011}. This, in turn, can be expressed as a Bloch wave of the form $\Pi_{h}=\R\{e^{iks}u_{k}(s)\}$, with $k$ a global wave number and $u_{k}=u_{k}(s)$ a function having the same period of the underlying lattice: i.e. $u_{k}(s)=u_{k}(s+a)$. The wave number $k$, analogous to the crystal momentum in one-dimensional solids, depends on the microstructure constant $K_{0}$ via the relation $k=\pm 1/a\,\arctan(\sqrt{4/\alpha^{2}-1})$, with $\alpha=2\cos(k_{-}b)\cos[k_{+}(a-b)]-(k_{+}/k_{-}+k_{-}/k_{+})\sin(k_{-}b)\sin[k_{+}(a-b)]$, with $k_{\pm}^{2}=(K_{0} \pm 1/R_{\pm}^{2})$ (see Ref.~\cite{Kollmitzer:2011} and Sec.~S6 in Ref.~\cite{SI}), corresponding to the typical band structure of Fig.~\ref{fig:metatubes}b. For $K_{0}$ values in the band gap, Eq.~\eqref{eq:shape} still admits a bounded solution of period $a$, which however does not convey {\em information} other than that already encoded in the metashell's architecture. Elastic waves with non-trivial dispersion relations and band gaps are not uncommon in solid mechanics (see e.g. Refs.~\cite{Kushwaha:1993,Hussein:2009}), but, to the best of the writer's knowledge, all known examples are limited to {\em acoustic} waves and are not, therefore, an exact analog of Bloch waves, which, on the other hand, consists of stationary configurations of the wave function obtained as solutions of the time-independent Schr\"odinger equation. Those discussed here provide then a first example of Bloch waves in classical solid mechanics and could possibly by employed in ``mechanical logic''~(see e.g. Ref.~\cite{Jiao:2023} for a recent perspective).

\noindent {\\[5pt]\bf Application to crystalline monolayers}\\
Before concluding, it is worth mentioning another realization of Eq.~\eqref{eq:pressure} with potential application to soft and biological matter: i.e. crystalline monolayers on curved interfaces. These are naturally found in colloidosomes~\cite{Dinsmore:2002,Bausch:2003,McGorty:2010}, interracially frozen emulsion droplets~\cite{Guttman:2016,GarciaAguilar:2021,Davidyan:2024}, as well as in simple multicellular organisms such as {\em Hydra}~\cite{Maroudas:2021,Maroudas:2024}. A classic approach to these systems, pioneered by Bowick {\em et al}. on phenomenological ground~\cite{Bowick:2000}, consists of treating topological defects, such as {\em dislocations} and {\em disclinations}, as fundamental degrees of freedom of an elastic theory, whose effective action is given by 
\begin{equation}\label{eq:bnt_model}
F=\frac{1}{2}\,Y\int {\rm d}A\,{\rm d}A'\,\mathcal{K}(\bm{r},\bm{r}')[\rho(\bm{r})-K(\bm{r})][\rho(\bm{r}')-K(\bm{r}')],
\end{equation}
where the kernel $\mathcal{K}$ is given by the biharmonic Green function -- i.e. $\nabla^{4}\mathcal{K}(\bm{r},\bm{r}')=\delta(\bm{r}-\bm{r}')$ -- and $\rho$ the so-called topological charge density expressing the angular deficit introduced by defects in the configuration of the displacement field: i.e. ${\rm d}({\rm Arg}\,\delta\bm{r})={\rm d}A\,\rho$~\cite{Bowick:2001,Bowick:2009}. Now, in flat crystalline monolayers, where $K=0$ and all components of the stress tensor can be expressed in terms of Airy stress function, Eq.~\eqref{eq:bnt_model} can be derived directly from Hooke's law~\cite{Seung:1988}. The same approach is however unavailable in the presence of a pre-existing Gaussian curvature, i.e. $K \ne 0$, and Eq.~\eqref{eq:bnt_model} is effectively obtained from a covariantization of the planar expression, followed by the crucial consideration that a local rotation of the displacement field is here compensated by the underlying Gaussian curvature~\cite{Bowick:2000,Bowick:2001,Bowick:2009}.

Now, with Eq.~\eqref{eq:pressure} in hand, taking $K'=\rho$ and $F=\int {d}A\,P^{2}/(2B)$ readily yields Eq.~\eqref{eq:bnt_model}, under the assumption of incompressibility:  i.e. $B\to\infty$. More interestingly, Eq.~\eqref{eq:pressure} provides a generalization of Eq.~\eqref{eq:bnt_model} to the compressible systems, where the bulk modulus is finite and the elastic deformations caused by the defects are either screened or enhanced by the Gaussian curvature of the substrate. 

\noindent {\\[5pt]\bf Conclusions}\\
In this article, I have explored the mechanical properties of elastic shells endowed of an internal microstructure and here referred to as {\em metashells}. Combining classic concepts of differential and integral geometry, I have shown that a configuration of the lateral pressure across the tangent plane of the shell is governed by a real-valued analog of the inhomogeneous Schr\"odinger equation, with the lateral pressure experienced by the microstructure in the role of the wave function. Using numerical simulations and analytical work, I illustrated two examples of phenomena with no counterpart in unstructured materials, but with some analogies in quantum systems. In the fine structure limit, that is when the internal structure is much smaller than the local radius of curvature of the metashell, the theory predicts the occurrence of elastic localization, by virtue of which deformations are prevented to diffuse away from their origin, thereby allowing the shell microstructure of smoothly adapt to the intrinsic geometry of the substrate. Leveraging on an analogy with scattering states in nearly free electrons, I showed that periodic metashells, obtained from the repetition of the same structural unit periodically in space, support Bloch waves, corresponding to stationary periodic deformations of the microstructure and characterized by a geometry-dependent band structure. If actuated by either natural or artificial stimuli, as in the shape-programmable magnetic materials developed in Ref.~\cite{Lum:2016}, these may find applications in the design of ``smart'' mechanical devices, where a single construction parameter -- i.e. the microstructure parameter $K_{0}$ -- maps into a specific {\em signal}, here encoded in the Floquet exponents of a periodic deformation and analogous to the crystal momentum in metallic solids. Furthermore, whereas the actual technological potential of similar devices is now difficult to foresee, there is in principle no fundamental hurdle preventing the manipulation of these signals by, e.g., the combination of different microstructures into diode- or triode-like terminals, with possible application to mechanical logic.

Aside from the context of metamaterials, the approach introduced here provides a possible route for modeling certain types of active and biological solids, where stresses originates at the scale of the individual building blocks. In these systems, one can imagine to parametrize {\em active strains} directly in terms of local variation of Gaussian curvature, possibly subject to global constraints reflecting the active nature of deformations, or coupled with additional internal fields accounting for the local configuration of the active subunits. Finally, when applied to crystalline interfaces, the framework delivered here provides a generalization of the elastic theory of interacting topological defects to compressible systems. 

\noindent {\\[5pt]\bf Acknowledgements}\\
I am indebted to Arthur Hernandez for many useful suggestions. This work is supported by the ERC-CoG grant HexaTissue and by Netherlands Organization for Scientific Research (NWO/OCW).

\input{main.bbl}

\end{document}


\title{Stretching theory of Hookean metashells\\Supplementary information}

\author{Luca Giomi}
\email{giomi@lorentz.leidenuniv.nl}
\affiliation{Instituut-Lorentz, Universiteit Leiden, P.O. Box 9506, 2300 RA Leiden, The Netherlands}

\maketitle

\section{\label{sec:pressure}Derivation of Eq.~(3)}

In this supplementary section, I provide a derivation of Eq.~(3), which represent the fundamental starting point to describe the configuration of the lateral pressure $P$, hence of the microstructure, of Hookean metashells. In Sec.~\ref{sec:pressure_preliminaries} we review some basic concept of surface geometry and establish some notational conventions, whereas Sec.~\ref{sec:pressure_derivation} is devoted to the derivation itself.

\subsection{\label{sec:pressure_preliminaries}Mathematical preliminaries and notation}

Let $\bm{r}=\bm{r}(x^{1},x^{2})$ be the position of a point on a generic surface embedded in $\mathbb{R}^{3}$ and parametrized in terms of the coordinates $\{x^{1},x^{2}\}$. Furthermore, let $\bm{g}_{i}=\partial_{i}\bm{r}$ be a  tangent vector along the $i-$th coordinate line, so that $g_{ij}=\bm{g}_{i}\cdot\bm{g}_{j}$ are the components of the the surface metric tensor and
\begin{equation}
\bm{N} = \frac{\bm{g}_{1}\times\bm{g}_{2}}{|\bm{g}_{1}\times\bm{g}_{2}|}\;,	
\end{equation}
the surface unit normal. The arc-length and area forms on the surface are given by
\begin{subequations}\label{eq:metric}
\begin{gather}
\D s^{2} = g_{ij}\D x^{i}\D x^{j}\;,\\[5pt]	
\D A = \sqrt{g}\,\D x^{i}\D x^{j}\;,
\end{gather}	
\end{subequations}
where we have called
\begin{equation}
g = g_{11}g_{22}-g_{12}g_{21}\;,
\end{equation}
the determinant of the metric tensor. Although in general not orthonormal, The tangent vectors $\{\bm{g}_{1},\bm{g}_{2}\}$ form a standard {\em covariant} basis for arbitrary vectors and tensors of any rank. The associated {\em contravariant} basis is obtained by means of the transformation $\bm{g}^{i}=g^{ij}\bm{g}_{j}$, so that, a generic tangent vector $\bm{v}$ is given by
\begin{equation}
\bm{v} = v^{i}\bm{g}_{i} = v_{i}\bm{g}^{i}\;,
\end{equation}
where $v_{i}=g_{ij}v^{j}$, while
\begin{equation}
\bm{g}^{i}\cdot\bm{g}_{j} = \delta^{i}_{j}\;,
\end{equation}
where $\delta^{i}_{j}$ is the Kronecker symbol: i.e. 
\begin{equation}
\delta^{1}_{1}=\delta^{2}_{2}=1\;,\qquad 
\delta^{1}_{2}=\delta^{2}_{1}=0\;.
\end{equation}
The surface curvature is routinely defined starting from the tensor of the second fundamental form, i.e.
\begin{equation}\label{eq:curvature_tensor}
b_{ij} = -\bm{g}_{i}\cdot\partial_{j}\bm{N}\;,
\end{equation}
whose trace and determinate respectively yield the mean curvature $H$ and Gaussian curvature $K$. That is
\begin{subequations}\label{eq:curvatures}
\begin{gather}
H = \frac{1}{2}\,g^{ij}b_{ij}\;,\\[5pt]
K = \frac{1}{2}\,\epsilon^{ik}\epsilon^{jl}b_{ij}b_{kl}\;,
\end{gather}	
\end{subequations}
where we have introduced the two-dimensional Levi-Civita tensor, whose components are
\begin{equation}\label{eq:levi_civita}
\epsilon_{11} = \epsilon_{22} = 0\;,\qquad 
\epsilon_{12} =-\epsilon_{21} = \sqrt{g}\;.
\end{equation}
The Levi-Civita tensor is characterized by the following normalization properties:
\begin{subequations}\label{eq:levi_civita_properties}
\begin{gather}
\epsilon^{ij} = \frac{\epsilon_{ij}}{g}\;,\\
\epsilon^{ik}\epsilon_{jk} = \delta^{i}_{j}\;,\\[5pt]
\epsilon^{ik}\epsilon_{jl} = \delta^{i}_{j}\delta^{k}_{l}-\delta^{i}_{l}\delta^{k}_{j}\;.
\end{gather}	
\end{subequations} 
The curvatures $H$ and $K$ and the tensors $g_{ij}$ and $b_{ij}$ are related, in turn, by the Cayley-Hamilton equation:
\begin{equation}\label{eq:cayley-hamilton}
b_{ik}b^{k}_{j} = 2Hb_{ij}-Kg_{ij}\;.
\end{equation}
Now, a crucial aspect of the derivation that will follow on Sec.~\ref{sec:pressure_derivation} revolves around the non-commutativity of covariant derivatives on surfaces endowed with a finite Gaussian curvature. Given a vector field $\bm{v}$, partial differentiation gives
\begin{equation}
\partial_{i}\bm{v} = \left(\nabla_{i}v^{j}\right)\bm{g}_{j} = \left(\nabla_{i}v_{j}\right)\bm{g}^{j}\;,
\end{equation}
where $\nabla_{i}$ indicate the covariant derivative, whose explicit form is given by
\begin{subequations}
\begin{gather}
\nabla_{i}v^{j} = \partial_{i}v^{j}+\Gamma_{ik}^{j}v^{k}\;, \\[5pt]
\nabla_{i}v_{j} = \partial_{i}v_{j}-\Gamma_{ij}^{k}v_{k}\;,
\end{gather}
\end{subequations}
respectively for the contravariant and covariant component of the vector field $\bm{v}$ and $\Gamma_{ij}^{k}=g^{kl}\Gamma_{ijl}$ is the Christoffel symbol of second kind, obtained upon raising one index of the Christoffel symbol of first kind. That is
\begin{equation}
\Gamma_{ijk} = \frac{1}{2}\left(\partial_{i}g_{kj}+\partial_{j}g_{ik}-\partial_{k}g_{ij}\right)\;.
\end{equation}
Unlike partial derivatives in flat space, covariant derivatives on intrinsically curved surfaces (i.e. surfaces where $K \ne 0$) only commute when acting on a scalar field. For vectorial and tensorial fields, on the other hand, it is possible to demonstrate the following expression for the commutator $[\nabla_{i},\nabla_{j}]=\nabla_{i}\nabla_{j}-\nabla_{j}\nabla_{i}$. That is
\begin{subequations}\label{eq:commutators}
\begin{gather}
\left[\nabla_{i},\nabla_{j}\right]v_{k} = R_{kji}^{l}v_{l}\;,\\[5pt]
\left[\nabla_{i},\nabla_{j}\right]T_{kl} = R_{kji}^{m}T_{ml}+R_{kji}^{m}T_{lm}\;,
\end{gather}
\end{subequations}
where $R^{i}_{jkl}$ is the mixed form of the Riemann curvature tensor. On surfaces, this can be expressed directly in terms of the metric tensor and the Gaussian curvature by means of the relation
\begin{equation}\label{eq:riemann_tensor}
R^{i}_{jkl} = K\left(\delta^{i}_{k}g_{jl}-g_{jk}\delta^{i}_{l}\right)\;.	
\end{equation}
Furthermore, using Eqs.~\eqref{eq:commutators} and \eqref{eq:riemann_tensor}, it is possible to prove the following additional relations for the commutators of the covariant divergence and Laplacian: i.e.
\begin{subequations}\label{eq:divergence_laplacian_commutators}
\begin{gather}
\left[\nabla^{i},\nabla_{j}\right]v_{i} = Kv_{j}\;,\\[5pt]
\left[\nabla_{i},\nabla^{2}\right]v_{j} = g_{ij}\nabla^{k}(Kv_{k})-\nabla_{j}(Kv_{i})\;,
\end{gather}
\end{subequations}
where we have introduced the Laplace-Beltrami operator, generalizing the standard Euclidean Laplacian. That is
\begin{equation}\label{eq:laplace_beltrami}
\nabla^{2} = \frac{1}{\sqrt{g}}\,\partial_{i}\left(\sqrt{g}\,g^{ij}\partial_{j}\right)\;.
\end{equation}
We conclude this section by recalling two useful identities involving the covariant derivatives of the tensors of the first and the second fundamental form: Ricci's Lemma and the equation of Mainardi-Codazzi. The former asserts
\begin{equation}
\nabla_{k}g_{ij} = 0\;,\qquad 	
\nabla_{k}g^{ij} = 0\;,\qquad 	
\nabla_{k}g = 0\;,	
\end{equation}
from which one also finds
\begin{equation}
\nabla_{k}\epsilon_{ij} = 0\;,\qquad
\nabla_{k}\epsilon^{ij} = 0\;.
\end{equation}
The equation of Mainardi-Codazzi, on the other hand, asserts that the rank$-3$ tensor $\nabla_{i}b_{jk}$ is totally symmetric. That is
\begin{equation}\label{eq:mainardi-codazzi}
\nabla_{i}b_{jk} = \nabla_{j}b_{ik}\;.
\end{equation}

\subsection{\label{sec:pressure_derivation}Derivation of the pressure equation}

Let us now imagine that the surface is subject to a generic small deformation, whose effect is to map a generic point of position $\bm{r}$ to $\bm{r}'=\bm{r}+\delta\bm{r}$, so that the local metric $\D s^{2}$, Eq.~(\ref{eq:metric}a), transforms into $\D s'^{2}=g_{ij}'\D x^{i} \D x^{j}$, with $g_{ij}'$ the metric tensor of the deformed surface. The strain tensor 
\begin{equation}
u_{ij} = \frac{g_{ij}'-g_{ij}}{2}\;,
\end{equation} 
is related to the stress tensor by Hooke's law, Eq.~(1), which could be alternatively expressed in the form
\begin{equation}\label{eq:hooke_law}
u_{ij} = \frac{1+\nu}{Y}\,\sigma_{ij}-\frac{\nu}{Y}\,\sigma_{k}^{k}g_{ij}\;,	
\end{equation}
with $\nu$ and $Y$ are the Poisson ratio and Young modulus respectively (see Table~1 in the main text). Furthermore, taking the trace of both side of Eq.~\eqref{eq:hooke_law} gives
\begin{equation}\label{eq:hooke_trace}
u_{i}^{i} = \frac{1-\nu}{Y}\,\sigma_{i}^{i}\;,
\end{equation}
whereas mechanical equilibrium requires the stress tensor to be divergence free. That is
\begin{equation}\label{eq:equilibrium}
\nabla_{j}\sigma^{ij} = 0 \;.
\end{equation}
Now, the strategy behind this derivation, consists of expressing the stress-strain incompatibility, defined as the double covariant curl of $u_{ij}$, to the linear variation of Gaussian curvature, that is 
\begin{equation}
\delta K = K'-K\;,
\end{equation}
where $K$ is the pre-existing Gaussian curvature of the shell and $K'$ that resulting from the elastic deformation. To this end, we parametrize the deformation $\delta\bm{r}$ in terms of tangential and normal displacements: i.e.
\begin{equation}\label{eq:deformation}
\delta\bm{r} = u^{i}\bm{g}_{i}+w\bm{N}\;,
\end{equation}
from which, using standard manipulations (see e.g. Ref.~[21]), one can explicit the strain tensor at the linear order in the form
\begin{equation}\label{eq:strain}
u_{ij} = \frac{1}{2}\left(\nabla_{i}u_{j}+\nabla_{j}u_{i}\right)-w b_{ij}\;.
\end{equation}
Contracting both sides of this equation and using Eq.~(\ref{eq:curvatures}a) gives, on the other hand, the following expression for the trace of the of the strain tensor:
\begin{equation}
u_{i}^{i} = \nabla_{j}u^{j}-2Hw\;.
\end{equation}
Next, contracting the left-hand side of Eq.~\eqref{eq:hooke_law} with $\epsilon^{ik}\epsilon^{jl}\nabla_{k}\nabla_{l}$, gives
\begin{subequations}\label{eq:incompatibility}
\begin{gather}
\epsilon^{ik}\epsilon^{jl}\nabla_{k}\nabla_{l}\sigma_{ij} = \nabla^{2}\sigma_{m}^{m}\;,\\[5pt]
\epsilon^{ik}\epsilon^{jl}\nabla_{k}\nabla_{l}\left(\sigma_{m}^{m}g_{ij}\right) = \nabla^{2}\sigma_{m}^{m}\;,\\[5pt] 	
\epsilon^{ik}\epsilon^{jl}\nabla_{k}\nabla_{l}u_{ij} = -\epsilon^{ik}\epsilon^{jl}b_{ij}\nabla_{k}\nabla_{l}w-\nabla_{m}\left(Ku^{m}\right)\;.
\end{gather}
\end{subequations}
Eqs.~(\ref{eq:incompatibility}a) and (\ref{eq:incompatibility}b) readily follow from the properties of the Levi-Civita tensor, i.e. Eqs.~\eqref{eq:levi_civita}, in the light of Eq.~\eqref{eq:equilibrium}. Eq.~(\ref{eq:incompatibility}c), on the other hand, will be derived later. The left-hand side of this equation, in turn, can be related with the linear variation of the Gaussian curvature. That is
\begin{equation}\label{eq:delta_k}
\delta K = \epsilon^{ik}\epsilon^{jl}b_{ij}\nabla_{k}\nabla_{l}w+\nabla_{m}(Ku^{m})-K\left(\nabla_{m}u^{m}-2Hw\right)\;,
\end{equation}
from which one readily obtains
\begin{equation}\label{eq:incompatibility_vs_delta_k}
\epsilon^{ik}\epsilon^{jl}\nabla_{k}\nabla_{l}u_{ij} = -\delta K - K\left(\nabla_{m}u^{m}-2Hw\right)\;.
\end{equation}
Combining Eqs.~\eqref{eq:hooke_law}, \eqref{eq:hooke_trace}, (\ref{eq:incompatibility}a), (\ref{eq:incompatibility}b) and \eqref{eq:incompatibility_vs_delta_k} then gives a single partial differential equation for the trace of the stress tensor. That is
\begin{equation}\label{eq:stress_pde}
\left[\nabla^{2}+\left(1-\nu\right)K\right]\sigma^{i}_{i} = - Y\delta K\;.	
\end{equation}
Finally, Eq.~(3) in the main text is then readily obtained upon setting $\sigma^{i}_{i}=-2P$ and $(1-\nu)=Y/(2B)$ (see Table~1).

To complete this derivation, we next demonstrate Eqs.~(\ref{eq:incompatibility}c) and \eqref{eq:delta_k}. The former can be conveniently accomplished starting from the expression of the strain tensor given in Eq.~\eqref{eq:strain}. Using the properties of the Levi-Civita tensor, Eq.~\eqref{eq:levi_civita}, gives, after some algebraic manipulations:
\begin{equation}
\epsilon^{ik}\epsilon^{jl}\nabla_{k}\nabla_{l} \left(\nabla_{i}v_{j}+\nabla_{j}v_{i}\right) 
= 2\nabla^{2}\left(\nabla^{m}u_{m}\right)-\nabla^{m}\left(\nabla^{2}u_{m}\right)-\nabla^{m}\nabla^{n}\nabla_{m}u_{n}\;.	
\end{equation}
Next, using Eq.~(\ref{eq:divergence_laplacian_commutators}a) to swap the order of $\nabla^{n}$ and $\nabla_{m}$ in the last term on the right-hand side of this equation and taking Eq.~(\ref{eq:divergence_laplacian_commutators}b) into account, yields
\begin{equation}
\epsilon^{ik}\epsilon^{jl}\nabla_{k}\nabla_{l} \left(\nabla_{i}v_{j}+\nabla_{j}v_{i}\right) = -\nabla_{m}\left(Ku^{m}\right)\;,	
\end{equation} 
which, together with the equation of Mainardi-Codazzi, Eq.~\eqref{eq:mainardi-codazzi}, readily gives Eq.~(\ref{eq:incompatibility}c). To demonstrate Eq.~\eqref{eq:delta_k}, on the other hand, it is convenient to use Eq.~(\ref{eq:curvatures}b), from which
\begin{equation}\label{eq:delta_gaussian}
\delta K 
= \epsilon^{ik}\left(\delta\epsilon^{jl}\right)b_{ij}b_{kl}
+ \epsilon^{ik}\epsilon^{jl}b_{ij}\left(\delta b_{kl}\right)\;.
\end{equation} 
The first term on the left-hand side of this equation can be computed starting from the definition of the Levi-Civita tensor, Eq.~\eqref{eq:levi_civita}. That is
\begin{equation}\label{eq:delta_epsilon}
\delta \epsilon^{ij} = -\left(\nabla_{k}u^{k}-2Hw\right)\epsilon^{ij}\;,
\end{equation}
where we used that the linear variation of the determinant of the metric: i.e.
\begin{equation}
\delta g = 2g\left(\nabla_{i}u^{i}-2Hw\right)\;.	
\end{equation}
Using Eq.~\eqref{eq:delta_epsilon} and Eq.~(\ref{eq:curvatures}b), allows us to express the first term on the right-hand side of Eq.~\eqref{eq:delta_gaussian} in the form
\begin{equation}\label{eq:delta_gaussian_1}
\epsilon^{ik}\left(\delta\epsilon^{jl}\right)b_{ij}b_{kl} = -2K\left(\nabla_{m}u^{m}-2Hw\right)\;.
\end{equation}
Similarly, to calculate the second term on the right-hand side of Eq.~\eqref{eq:delta_gaussian}, we recall that, at the linear order, the variation of the curvature tensor is given by Ref.~[25]:
\begin{equation}\label{eq:delta_b}
\delta b_{ij} = b_{ik}\nabla_{j}u^{k}+b_{jk}\nabla_{i}u^{k}+u^{k}\nabla_{k}b_{ij}+\nabla_{i}\nabla_{j}w-b_{ik}b^{k}_{k}w\;.	
\end{equation}
Each of these five terms must be know contracted with $\epsilon^{ik}\epsilon^{jk}b_{ij}$ in order to compute the curvature variation. Using the identities reviewed in Sec.~\ref{sec:pressure_preliminaries}, and in particular Eqs.~\eqref{eq:levi_civita_properties} and \eqref{eq:cayley-hamilton}, allows one to obtain the following useful expressions
\begin{subequations}
\begin{gather}
\epsilon^{ik}\epsilon^{jl}b_{ij}b_{km}\nabla_{l}u^{m} = \epsilon^{ik}\epsilon^{jl}b_{ij}b_{lm}\nabla_{k}u^{m} = K\nabla_{n}u^{n}\;,\\[5pt]
\epsilon^{ik}\epsilon^{jl}b_{ij}u^{m}\nabla_{m}b_{kl} = u^{n}\nabla_{n}K\;,\\[5pt]
\epsilon^{ik}\epsilon^{jl}b_{ij}b_{km}b^{m}_{l} = 2HK\;,
\end{gather}	
\end{subequations} 
from which one can readily derive a concise expression for the second term on the right-hand side of Eq.~\eqref{eq:delta_gaussian}. That is
\begin{equation}\label{eq:delta_gaussian_2}
\epsilon^{ik}\epsilon^{jl}b_{ij}\left(\delta b_{kl}\right) = \epsilon^{ik}\epsilon^{jl}b_{ij}\nabla_{k}\nabla_{l}w + \nabla_{m}\left(Ku^{m}\right)+K\left(\nabla_{m}u^{m}-2Hw\right)\;.
\end{equation}
Finally, adding Eqs.~\eqref{eq:delta_gaussian_1} and ~\eqref{eq:delta_gaussian_2} gives Eq.~\eqref{eq:delta_k} for the linear variation of the Gaussian curvature.

To conclude this supplementary section we show that from Eq.~(3), or equivalently from Eq.~\eqref{eq:stress_pde}, it is possible to derive the standard mechanical equilibrium condition of spherical shells. In this case, $K=1/R^{2}$, with $R$ the radius of the undeformed shell. The property of the Gaussian curvature being uniform throughout the surface allows one to parametrize the stress tensor in terms of an Airy stress function $\Phi$, so that
\begin{equation}\label{eq:airy}
\sigma^{ij} = \left(\epsilon^{ik}\epsilon^{jl}\nabla_{k}\nabla_{l}+Kg^{ij}\right)\Phi\;.	
\end{equation}
As in the case of plates, where $K=0$, Eq.~\eqref{eq:airy} automatically guarantees mechanical equilibrium, i.e. Eq.~\eqref{eq:equilibrium}, by rendering the stress tensor divergence-free. Explicitly:
\begin{equation} 
\nabla_{j}\sigma^{ij} = \Phi\nabla^{i}K = 0\;,
\end{equation}
where the first equality can be derived using Eqs.~(\ref{eq:commutators}a) and \eqref{eq:riemann_tensor}, with $v_{k}=\nabla_{k}\Phi$, while the second evidently holds only for surfaces with constant Gaussian curvature. Contracting both sides of Eq.~\eqref{eq:airy} then gives
\begin{equation}\label{eq:pressure_sphere}
\sigma^{i}_{i} = \left(\nabla^{2}+\frac{2}{R^{2}}\right)\Phi\;.
\end{equation}
The corresponding Gaussian curvature variation, in turn, can be made explicit by taking into account that, on an outward-oriented sphere, $b_{ij}=-g_{ij}/R$ and $H=-1/R$. Thus, from Eq.~\eqref{eq:delta_k}, we find
\begin{equation}\label{eq:delta_k_sphere}
\delta K 
= \epsilon^{ik}\epsilon^{jl}\nabla_{k}\nabla_{l}w+2HKw 
=-\left(\nabla^{2}+\frac{2}{R^{2}}\right)\frac{w}{R}\;.
\end{equation}
Finally, replacing Eq.~\eqref{eq:pressure_sphere} and \eqref{eq:delta_k_sphere} in Eq.~\eqref{eq:stress_pde} gives, after simple manipulations
\begin{equation}\label{eq:spherical_shells}
\left(\nabla^{2}+\frac{2}{R^{2}}\right)\left[\left(\nabla^{2}+\frac{1-\nu}{R^{2}}\right)\frac{\Phi}{Y}-\frac{w}{R}\right] = 0\;,
\end{equation}
thus recovering the classic equation describing the distribution of in-plane stresses in spherical shells subject to a normal displacement of magnitude $w$ (see e.g. Ref.~[25]).

\section{\label{sec:stress}Derivation of Eq.~(5)}

As explained in the main text, the strategy behind the derivation of Eq.~(5) is built upon the assumption that $K'$ is a motion invariant {\em valuation} of a patch of the metashell where the lateral pressure is approximatively uniform. Hadwiger's characterization theorem allows then to express $K'$ in terms of the area $\mathcal{A}'$, perimeter $\mathcal{L}'$ and Euler's characteristic $\chi'$ of the {\em deformed} patch. That is
\begin{equation}\label{eq:hadwiger}
K' = \alpha_{0}\mathcal{A}'+\alpha_{1}\mathcal{L}'+\alpha_{2}\chi'\;,	
\end{equation}
where $\alpha_{0}$, $\alpha_{1}$ and $\alpha_{2}$ constants. Because in this construction the shape of a unit cell is assumed to depart from the initial one  only in response to the adaptation of the cell to the substrate's intrinsic geometry, $K'=0$ when $\mathcal{A}'$, $\mathcal{L}'$ and $\mathcal{\chi}'$ equate those of the original shape: i.e. 
\begin{equation}
\alpha_{0}\mathcal{A}+\alpha_{1}\mathcal{L}+\alpha_{2}\mathcal{\chi}=0\;.
\end{equation}
Furthermore, as the type of deformations considered here do not alter the Euler characteristic, $\alpha_{2}\chi'=\alpha_{2}\chi=-\alpha_{0}\mathcal{A}-\alpha_{1}\mathcal{L}$. This allows to express the right-hand side of Eq.~\eqref{eq:hadwiger} in terms of the variations $\delta\mathcal{A}=\mathcal{A}'-\mathcal{A}$ and $\delta\mathcal{L}=\mathcal{L}'-\mathcal{L}$ of the area and perimeter of the patch. That is
\begin{equation}\label{eq:hadwiger_delta}
K' = \alpha_{0}\delta\mathcal{A}+\alpha_{1}\delta\mathcal{L}\;.
\end{equation}
To make progress, we parametrize the shape variation in terms of a small displacement of the boundary of the patch along the tangent-normal direction $\bm{\nu}=\bm{t}\times\bm{N}$: i.e. $\bm{u}=\epsilon\bm{\nu}$. This gives (see e.g. Ref.~[\href{https://doi.org/10.1039/C3SM50484K}{29}])
\begin{subequations}\label{eq:area_perimeter_variations}
\begin{gather}
\delta\mathcal{A} = \oint {\rm d}s\,\epsilon = \mathcal{L}\epsilon\;,\\
\delta\mathcal{L} = \oint {\rm d}s\,\kappa\epsilon = \Theta\epsilon\;,
\end{gather}
\end{subequations}
where we have used the assumption that $\epsilon$ uniform within the patch and in particular along the boundary. The constant $\Theta$ is the angle the tangent vector $\bm{t}$ rotates in one loop along the boundary and depends on the regularity of curve and can be smaller, equal or larger than $2\pi$. The first scenario correspond to the case of a piecewise continuous boundary, where the tangent vector undergoes discontinuous rotations while crossing a discrete number of ``kinks''. The second and third scenarios, on the other hand, correspond respectively to the case of simple and self-intersecting smooth curves. In these cases $\Theta=2\pi k$, with $k$ an integer enumerating the number of turns. The tangent-normal displacement $\epsilon$ can finally be expressed directly in terms of the dimensionless pressure $\Pi$. To this end, we take the trace of Eq.~(1) to obtain
\begin{equation}
\mathcal{P} = - B\left(\nabla\cdot\bm{u}\right)\;.
\end{equation}
Then, averaging both sides of this equation over the interior of the cell and using the divergence theorem gives
\begin{equation}\label{eq:displacement}
\Pi = -\int \frac{{\rm d}A}{\mathcal{A}}\,\left(\nabla\cdot\bm{u}\right) = -\oint \frac{{\rm d}s}{\mathcal{A}}\,\epsilon = -\frac{\mathcal{L}}{\mathcal{A}}\,\epsilon\;.
\end{equation}
where we have set $\Pi=(1/\mathcal{A})\int {\rm d}A\,(\mathcal{P}/B)$. Finally, replacing Eqs.~\eqref{eq:area_perimeter_variations} and \eqref{eq:displacement} in Eq.~\eqref{eq:hadwiger_delta}, gives
\begin{equation}\label{eq:keff}
K' = - \left(\alpha_{0}+\alpha_{1}\,\frac{\Theta}{\mathcal{L}}\right)\mathcal{A}\Pi\;,
\end{equation}
hence Eq.~(5) in the main text, with $K_{0}=(\alpha_{0}+\alpha_{1}\Theta/\mathcal{L})\mathcal{A}$.

\section{Estimate of $K_{0}$ for positionally constrained geodesic polygons}

\begin{figure}
\centering
\includegraphics[width=\textwidth]{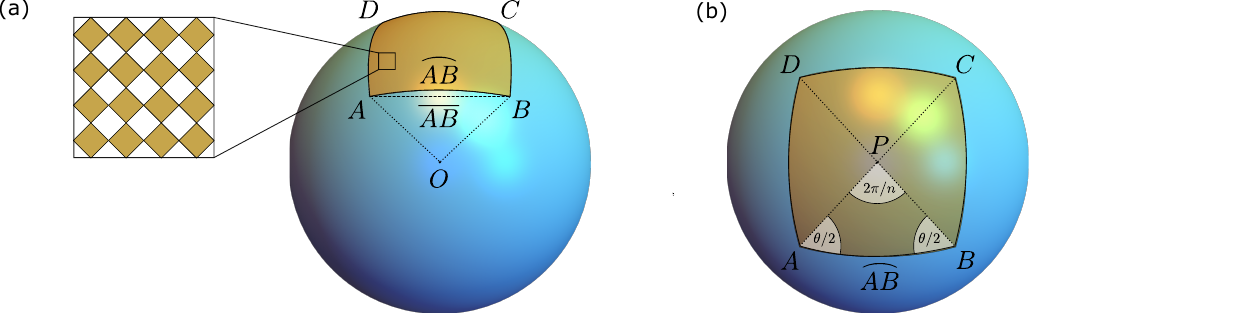}
\caption{\label{fig:geodesic_polygon}(a) Example of regular geodesic polygon on a sphere centered at the point $O$. (b) Top view of the same polygon, which, by virtue of its regularity, can be partitioned into $n$ identical isosceles triangles, with $n$ the number of sides of the polygon ($n=4$ in this example).}
\end{figure}

In the supplementary section, we discuss how the material constant $K_{0}$, introduced in Eq.~(3), can be computed upon assuming a specific deformation model for the shell microstructure. For sake of simplicity, we consider a toy model in which a patch deformed metashell consists of a regular $n-$sided {\em geodesic} polygons, whose vertices preserve their position in the three-dimensional ambient space. This positional constraint renders this construction independent of the material parameters. In the flat configuration, the patch has perimeter $\mathcal{L}$ and area $\mathcal{A}=\mathcal{L}^{2}/(4n)\,\cot(\pi/n)$. If the polygon is now laid on a substrate having local Gaussian curvature $K'$, with the latter assumed roughly constant across its length, it can be shown that its perimeter and area change as follows:
\begin{subequations}\label{eq:lp_ap}
\begin{gather}
\mathcal{L}' \approx \mathcal{L}\left[1+\frac{K'}{6}\left(\frac{\mathcal{L}}{2n}\right)^{2}\right]\;,\\[5pt]
\mathcal{A}' \approx \mathcal{A}\left[1+K'\left(\frac{\mathcal{L}}{2n}\right)^{2}\right]\;,
\end{gather}
\end{subequations}
where the approximation holds at the quadratic order in $K\mathcal{L}^{2}$, with a correction of order $\mathcal{O}(K^{2}\mathcal{L}^{2})$. Combining Eq.~(\ref{eq:lp_ap}a) and Eq.~(\ref{eq:lp_ap}b) and solving the resulting single linear equation with respect to $K'$, readily yields and expression of the effective Gaussian curvature of the form given in Eq.~\eqref{eq:hadwiger_delta}, from which one can identify the coefficients $\alpha_{0}$ and $\alpha_{1}$. That is
\begin{equation}
K' 
= \underbrace{\frac{2n^{2}}{\mathcal{A}\mathcal{L}^{2}}}_{\alpha_{0}}\,\left(\mathcal{A}'-\mathcal{A}\right)	
+ \underbrace{\frac{12n^{2}}{\mathcal{L}^{3}}}_{\alpha_{1}}\,\left(\mathcal{L}'-\mathcal{L}\right)\;,
\end{equation}
from which, using Eq.~\eqref{eq:keff}, one finds
\begin{equation}
K_{0} = \frac{2n^{2}}{\mathcal{L}^{2}}\,\left(1+6\Theta\,\frac{\mathcal{A}}{\mathcal{L}^{2}}\right)\;.
\end{equation}
To complete this demonstration we now derive Eqs.~\eqref{eq:lp_ap}. To fix ideas one can consider regular $n-$sided polygon on a sphere of radius $R$, as illustrated in Fig.~\ref{fig:geodesic_polygon} in the case $n=4$. Up to quadratic order in $\mathcal{L}/R$, the latter serves as approximation of any local neighbourhood featuring a positive Gaussian curvature $K'=1/R^{2}$. To derive Eq.~(\ref{eq:lp_ap}a), it is sufficient to recall the relation between the geodesic and chordal distance between, say, the vertices $A$ and $B$ on the sphere (see Fig.~\ref{fig:geodesic_polygon}a). That is
\begin{equation}\label{eq:arc_chord_spherical}
|\arc{AB}| = 2R\arcsin\left(\frac{|\overline{AB}|}{2R}\right)\;,
\end{equation}
where the curved (straight) bar indicates the geodesic (chord) connecting $A$ and $B$ and $|\cdots|$ its length. As all edges have equal length in a regular polygons, we find
\begin{equation}\label{eq:lp_vs_l}
\mathcal{L}' = 2nR\arcsin\left(\frac{\mathcal{L}}{2nR}\right)\;.
\end{equation}
To compute the area of the polygon, on the other hand, we make use of the generic expression
\begin{equation}\label{eq:spherical_area}
\mathcal{A}' = R^{2}\left[\sum_{i=1}^{n}\theta_{i}-(n-2)\pi\right]\;.	
\end{equation}
where $\theta_{i}$, with $i=1,\,2\ldots\,n$, are the {\em internal} angles of the polygon and the expression can be derived, for instance, by applying the Gauss-Bonnet theorem to a closed domain $\Omega$, whose boundary $\partial\Omega$ features an arbitrary number of ``kinks'' (i.e. points where the tangent vector undergoes a discontinuous rotation). That is: 
\begin{equation}\label{eq:gauss_bonnet}
\int_{\Omega} {\rm d}A\,K +	\oint_{\partial\Omega} {\rm d}s\,\kappa_{g} + \sum_{i}\phi_{i} = 2\pi\;,
\end{equation}
where $\kappa_{g}$ is the geodesic curvature of the domain's boundary $\partial\Omega$ and $\phi_{i}=\pi-\theta_{i}$ are the {\em external} angles at the kinks. Since $K$ is constant on the sphere and $\kappa_{g}=0$ along any geodesic, Eq.~\eqref{eq:gauss_bonnet} readily gives Eq.~\eqref{eq:spherical_area} when applied to a $n-$sided and not necessarily regular geodesic polygon. Since in a regular polygon all internal angles are equal, $\theta_{i}=\theta$ with $i=1,\,2\ldots\,n$, and Eq.~\eqref{eq:spherical_area} can be cast in the form
\begin{equation}\label{eq:theta_vs_a}
\cos\theta = \cos\left(\frac{\mathcal{A}'}{nR^{2}}+\frac{n-2}{n}\,\pi\right)\;.
\end{equation}
The left-hand side of this equation can be, in turn, expressed in terms of the perimeter $\mathcal{L}'$ by means of the so called law of cosines for spherical triangles~[\href{https://www.gutenberg.org/ebooks/19770}{51}]. When applied to the spherical triangle $\triangle ABP$ in Fig.~\ref{fig:geodesic_polygon}, this demands
\begin{equation}\label{eq:law_of_cosines}
\cos(\angle{P})=-\cos(\angle{A})\cos(\angle{B})+\sin(\angle{A})\sin(\angle{B})\cos\left(\frac{|\arc{AB}|}{R}\right)\;,	
\end{equation}
where $\angle{V}$ indicates the angle associated with the generic vertex $V$ of the triangle. In this case, $\angle{A}=\angle{B}=\theta/2$, $\angle{P}=2\pi/n$ and $|\arc{AB}|=\mathcal{L}/n$, thus, using Eq.~\eqref{eq:law_of_cosines}, one obtains
\begin{equation}\label{eq:costheta}
\cos\theta = 1-2\,\frac{1+\cos\left(\frac{2\pi}{n}\right)}{1+\cos\left(\frac{\mathcal{L}}{n}\right)}\;.		
\end{equation}
Next, using Eq.~\eqref{eq:costheta} and expanding the right-hand side of Eq.~\eqref{eq:theta_vs_a} at the linear order in $\mathcal{A}'/R^{2}$, gives
\begin{equation}
\cos\left(\frac{2\pi}{n}\right)+\sin\left(\frac{2\pi}{n}\right)\frac{\mathcal{A}'}{nR^{2}} 
\approx 2\,\frac{1+\cos\left(\frac{2\pi}{n}\right)}{1+\cos\left(\frac{\mathcal{L}}{n}\right)}-1\;,	
\end{equation}
which, after some algebraic manipulations finally yields
\begin{equation}\label{eq:ap_vs_l}
\mathcal{A}' 
\approx nR^{2}\tan^{2}\left(\frac{\mathcal{L}'}{2nR}\right)\cot\left(\frac{\pi}{n}\right)
= \frac{\mathcal{L}^{2}}{4n\left[1-\left(\frac{\mathcal{L}}{2nR}\right)^{2}\right]}\,\cot\left(\frac{\pi}{n}\right)\;,
\end{equation}
where the second equality is obtained by virtue of Eq.~\eqref{eq:lp_vs_l}. Lastly, expanding Eq.~\eqref{eq:lp_vs_l} and \eqref{eq:ap_vs_l} at the quadratic order in $\mathcal{L}/R$ gives
\begin{subequations}\label{eq:lp_ap_spherical}
\begin{gather}
\mathcal{L}' \approx \mathcal{L}\left[1+\frac{1}{6}\left(\frac{\mathcal{L}}{2nR}\right)^{2}\right]\;,\\[5pt]
\mathcal{A}' \approx \mathcal{A}\left[1+\left(\frac{\mathcal{L}}{2nR}\right)^{2}\right]\;.
\end{gather}
\end{subequations}
The same derivation can be repeated for a $n-$sided regular geodesic polygon on a pseudospherical surface having constant negative Gaussian curvature $K'=-1/R^{2}$. An analog of Eq.~\eqref{eq:arc_chord_spherical}, expressing the relation between geodesic and chordal distance, is given, in this case, by
\begin{equation}\label{eq:arc_chord_pseudospherical}
|\arc{AB}| = 2R\sinh\left(\frac{|\overline{AB}|}{R}\right)\;,	
\end{equation}
whereas the area of the polygon is given by
\begin{equation}
\mathcal{A}' = -R^{2}\left[\sum_{i=1}^{n}\theta_{i}-(n-2)\pi\right]\;,
\end{equation}
as one can readily verify again from the Gauss-Bonnet theorem: i.e. Eq.~\eqref{eq:gauss_bonnet}. Then, following the same algebraic manipulations used in the spherical case, one readily finds
\begin{equation}
\mathcal{A}' 
= -nR^{2} \tan^{2}\left(\frac{\mathcal{L}'}{2nR}\right)\cot\left(\frac{\pi}{n}\right)
= \frac{\mathcal{L}^{2}}{4n\left[1+\left(\frac{\mathcal{L}}{2nR}\right)^{2}\right]}\,\cot\left(\frac{\pi}{n}\right)\;.
\end{equation}
Finally, expanding this and Eq.~\eqref{eq:arc_chord_pseudospherical} at the quadratic order in $\mathcal{L}/R$ gives
\begin{subequations}\label{eq:lp_ap_pseudospherical}
\begin{gather}
\mathcal{L}' \approx \mathcal{L}\left[1-\frac{1}{6}\left(\frac{\mathcal{L}}{2nR}\right)^{2}\right]\;,\\[5pt]
\mathcal{A}' \approx \mathcal{A}\left[1-\left(\frac{\mathcal{L}}{2nR}\right)^{2}\right]\;.
\end{gather}
\end{subequations}
Comparing Eqs.~\eqref{eq:arc_chord_spherical} and \eqref{eq:arc_chord_pseudospherical} it is now straightforward to notice that these expressions differ only for the sign of the quadratic correction. Thus, setting $K'=1/R^{2}$ in the former and $K'=-1/R^{2}$ in the latter, allows one to express $\mathcal{L}'$ and $\mathcal{A}'$ as in Eqs.~\eqref{eq:lp_ap}, thereby completing our derivation.

\section{\label{sec:numerics}Derivation of Eq.~(6)}

Thanks to the linearity of Eq.~(5), it always possible to express its solution as a linear combination of a particular solution, $\Pi_{p}$, and the solution of the associated homogeneous equation, $\Pi_{h}$. Thus
\begin{equation}\label{eq:exact_sol}
\Pi = \Pi_{p}+\Pi_{h}\;.
\end{equation}
In the following, we will see how an exact particular solution can be constructed as an infinite sum of even-valued powers of the covariant Laplacian.  To make progress, it is convenient to rescale $K$ by $K_{0}$ and the position vector $\bm{r}$ by the system size $L$. For the three surfaces of revolution considered here, this equates the range of the geodesic latitude $s$ introduced in the main text: i.e. $-L/2\leq s \leq L/2$. More generically, $L$ represents the shortest distance between any two boundaries of the surface or between a point and itself in the case of closed surfaces. Eq.~(5) becomes then
\begin{equation}\label{eq:torus_dimensionsless}
\epsilon \partial^{2}\Pi_{p}+(1+\kappa)\Pi_{p}+\kappa = 0\;,	
\end{equation}
where $\kappa=K/K_{0}$ and $\partial^{2}=L^{2}\nabla^{2}$. The parameter $\epsilon$, on the other hand, is given by the ratio
\begin{equation}
\epsilon = \frac{\beta}{K_{0}L^{2}}\;,	
\end{equation}
and is, by construction, much smaller than one: i.e. $\epsilon \ll 1$. Taking advantage of the latter condition, one can look for a solution of the form
\begin{equation}\label{eq:perturbative}
\Pi_{p} = \Pi_{0}+\epsilon\Pi_{1}+\epsilon^{2}\Pi_{2}+\cdots\;
\end{equation}
Replacing this in Eq.~\eqref{eq:torus_dimensionsless} and equating the coefficients of the same powers of $\epsilon$, yields the following set of recursive algebraic equations
\begin{subequations}
\begin{gather}\label{eq:recursive}
(1+\kappa)\Pi_{0} + \kappa = 0\;,\\[5pt]	
(1+\kappa)\Pi_{n} + \partial^{2}\Pi_{n-1} = 0\;,\qquad n \ge 1\;,
\end{gather}
\end{subequations}
from which one readily obtains
\begin{align*}	
&\Pi_{0} = -\frac{\kappa}{1+\kappa}\;,\\[2.5pt]
&\Pi_{1} = -\frac{1}{1+\kappa}\,\partial^{2}\left(-\frac{\kappa}{1+\kappa}\right)\;,\\[2.5pt]
&\Pi_{2} = -\frac{1}{1+\kappa}\,\partial^{2}\left[-\frac{1}{1+\kappa}\,\partial^{2}\left(-\frac{\kappa}{1+\kappa}\right)\right]\;,\\[2.5pt]
&\cdots
\end{align*}
Using this and Eq.~\eqref{eq:perturbative} and restoring the original dimensions finally gives
\begin{equation}\label{eq:pi_p}
\Pi_{p} = \sum_{n=0}^{\infty}(-1)^{n+1}\mathscr{D}^{2n}\left(\frac{K}{K_{0}+K}\right)\;,	
\end{equation}
where we have introduced the differential operator
\begin{equation}
\mathscr{D}^{2}(\cdots) = \frac{\beta}{K_{0}+K}\,\nabla^{2}(\cdots)\;,
\end{equation}
as well as its recursive form
\begin{equation}
\mathscr{D}^{2n}(\cdots) = \underbrace{\mathscr{D}^{2}\circ \mathscr{D}^{2} \circ \cdots \circ \mathscr{D}^{2}}_{n\,{\rm times}}(\cdots)\;,
\end{equation}
thereby completing the derivation of Eq.~(6).

\section{\label{sec:analytical} Analytical solution of Eq.~(5) for barrels and toroidal metashells}

\begin{figure}
\centering
\includegraphics[width=\textwidth]{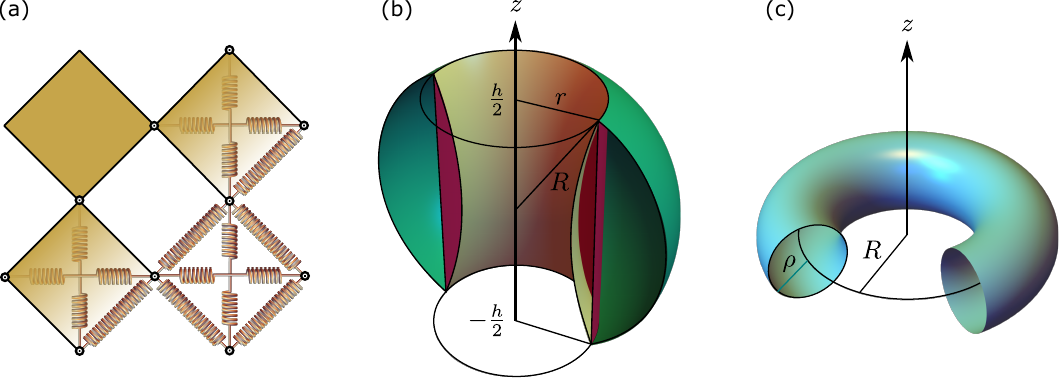}
\caption{\label{fig:numerics}(a) Cross-section of spherical (green) and pesudospherical (yellow) barrels having the same hight, boundaries and Gaussian curvature. Both configurations are constructing starting from an auxiliary cylindrical kirigami shell (red) and minimizing the total elastic energy, Eq.~\eqref{eq:spring_energy}. (b) Cross-section of a torus with radii $\rho<R$. (c) A unit cell of the kirigami metashell discussed in the main text. In numerical simulations, the unit cell is modelled by means of a network of Hookean springs and rigidified by linking the vertices across the diagonals.}
\end{figure}

Eq.~\eqref{eq:pi_p} [Eq.~(6) in the main text] can be readily employed to calculate the particular solution of Eq.~(5) for the three surfaces of revolution considered here. In the following we will complement this with the solution $\Pi_{h}$ of the associated homogeneous equation, that is
\begin{equation}\label{eq:homogeneous}
\beta \nabla^{2}\Pi_{h}+(K_{0}+K)\Pi_{h}=0\;.
\end{equation}
To make progress, we parameterize the surfaces in terms of the geodesic latitude $-L/2\le s \le L/2$, that is the distance from the equator measured along any of the surfaces' meridians, and the azimuthal angle $0\le \phi < 2\pi$. In these variables, spherical barrels ($\mathbb{S}^{2}$) and pseudospherical barrales ($\mathbb{\Sigma}^{2}$) take the following parametric form
\begin{equation}\label{eq:barrels_parametrization}
\mathbb{S}^{2}:\quad
\left\{
\begin{array}{l}
x = R\cos\left(\frac{s}{R}\right)\cos\phi\\[5pt]
y = R\cos\left(\frac{s}{R}\right)\sin\phi\\[5pt]
z = R\sin\left(\frac{s}{R}\right)
\end{array}
\right.\;,
\qquad
\mathbb{\Sigma}^{2}:\quad
\left\{
\begin{array}{l}
x = \rho\cosh\left(\frac{s}{R}\right)\cos\phi\\[5pt]
y = \rho\cosh\left(\frac{s}{R}\right)\sin\phi\\[3pt]
z =-iR E\left(\frac{is}{R}\Big|-\frac{\rho^{2}}{R^{2}}\right)\;,
\end{array}
\right.\;,
\end{equation}
with $R$ the radius of curvature (see Fig.~\ref{fig:numerics}a), $E(\phi|m)=\int {\rm d}t\,(1-m\sin^{2} t)^{1/2}$, with $|m|\le 1$, the incomplete elliptic integral of second kind [52] and $\rho$ a positive constant depending on the system size. In both cases, the radius of curvature $R$ can be related with the width and height of the surfaces. For spherical barrels, in particular, one readily finds
\begin{equation}
R = \sqrt{r^{2}+\left(\frac{h}{2}\right)^{2}}\;,	
\end{equation}
where $r$ is the radius of the circular rim and $h$ the height. In the case of pseudospherical barrels, on the other hand, fixing $r$ and $h$ allows multiple values of $R$ and $\rho$, which, in turn, must satisfy the equation
\begin{equation}
\frac{h}{2R} + iE\left(i\arccosh\frac{r}{\rho}\bigg|-\frac{\rho^{2}}{R^{2}}\right) = 0\;,
\end{equation}
whose solutions are subject to the constraints
\begin{subequations}
\begin{gather}
\rho \le R\;,\\[5pt]
\rho^{2} \le r^{2} \le \rho^{2}+R^{2}\;,
\end{gather}
\end{subequations}
in order for the coordinates to be real-valued. The Gaussian curvature, finally, is uniform across both surfaces and given by 
\begin{equation}
K = \pm \frac{1}{R^{2}}\;. 
\end{equation}
The embedded torus ($\mathbb{T}^{2}$) can be similarly parametrized in terms of the distance $s$ from the external equator and the azimuthal angle $\phi$ (see Fig.~\ref{fig:numerics}b). This gives
\begin{equation} 
\mathbb{T}^{2}:\quad 
\left\{
\begin{array}{l}
x = \left[R+\rho\cos\left(\frac{s}{\rho}\right)\right]\cos\phi\\[5pt]
y = \left[R+\rho\cos\left(\frac{s}{\rho}\right)\right]\sin\phi\\[5pt]
z = \rho\sin\left(\frac{s}{\rho}\right)\;.
\end{array}
\right.\;,
\end{equation}
where $R>\rho$ now represent the two radii of the torus. Unlike barrels, toroidal substrates feature a position-dependent Gaussian curvature, given by
\begin{equation}
K = \frac{\cos\left(\frac{s}{\rho}\right)}{\rho\left[R+\rho\cos\left(\frac{s}{\rho}\right)\right]}\;.	
\end{equation}
This is positive in the exterior of the torus, where $-L/4\le s \le L/4$, with $L=2\pi\rho$ the cross-sectional circumference, negative in the interior and maximal in magnitude along the internal equator, where $s=\pm L/2$ and $K=-1/[\rho(R-\rho)]$. 

Now, in the case of spherical and pseudospherical barrels, the only non-vanishing term of the sum in Eq.~\eqref{eq:pi_p} [Eq.~(6) in the main text] corresponds to $n=0$ and the particular solution $\Pi_{p}$ takes the simple form
\begin{equation}
\Pi_{p} = -\frac{1}{1\pm K_{0}R^{2}}\;. 	
\end{equation}
To solve the associated homogeneous equation, we first take advantage of the rotational symmetry of the surfaces to reduce Eq.~\eqref{eq:homogeneous} to a simpler ordinary differential equation. To this end, we first use Eqs.~\eqref{eq:barrels_parametrization} to calculate the coefficients of the metric tensor $g_{ij}$ with $\{i,j\}=\{s,\phi\}$. This gives
\begin{equation}\label{eq:barrels_metric}
\mathbb{S}^{2}:\quad
\left\{
\begin{array}{l}
g_{ss} = 1\\[5pt]
g_{\phi\phi} = R^{2}\cos^{2}\left(\frac{s}{R}\right)\\[5pt]
g_{s\phi} = g_{\phi s} = 0
\end{array}\;,
\right.
\qquad 
\mathbb{\Sigma}^{2}:\quad
\left\{
\begin{array}{l}
g_{ss} = 1\\[5pt]
g_{\phi\phi} = \rho^{2}\cosh^{2}\left(\frac{s}{R}\right)\\[5pt]
g_{s\phi} = g_{\phi s} = 0
\end{array}
\right.\;,
\end{equation}
from which, using Eq.~\eqref{eq:laplace_beltrami}, one can reduce Eq.~\eqref{eq:homogeneous} to the form
\begin{subequations}\label{eq:barrels_ode}
\begin{align}
\mathbb{S}^{2}:
&\qquad\partial_{s}^{2}\Pi_{h}-\frac{1}{R}\,\tan\left(\frac{s}{R}\right)\,\partial_{s}\Pi_{h}+k_{+}^{2}\Pi_{h}=0\;,\\[7pt]
\mathbb{\Sigma}^{2}: 
&\qquad\partial_{s}^{2}\Pi_{h}+\frac{1}{R}\,\tanh\left(\frac{s}{R}\right)\,\partial_{s}\Pi_{h}+k_{-}^{2}\Pi_{h}=0\;,
\end{align}
\end{subequations}
where the double prime denotes a second derivative with respect to $s$ and $k_{\pm}^{2}=(K_{0}\pm 1/R^{2})/\beta$. Eqs.~\eqref{eq:barrels_ode} can now be solved in terms of the Legendre functions and associated Legendre functions~[52] to give
\begin{subequations}\label{eq:barrels_sol}
\begin{align}
\mathbb{S}^{2}:
&\qquad\Pi_{h} = c_{1}P_{\lambda_{+}-\frac{1}{2}}\left[\sin\left(\frac{s}{R}\right)\right]+c_{2}Q_{\lambda_{+}-\frac{1}{2}}\left[\sin\left(\frac{s}{R}\right)\right]\;,\\[7pt]
\mathbb{\Sigma}^{2}: 
&\qquad\Pi_{h} = \left[1-\tanh^{2}\left(\frac{s}{R}\right)\right]^{\frac{1}{4}}\left\{ c_{1}P_{\lambda_{-}}^{-\frac{1}{2}}\left[\tanh\left(\frac{s}{R}\right)\right]+c_{2}Q_{\lambda_{-}}^{-\frac{1}{2}}\left[\tanh\left(\frac{s}{R}\right)\right]\right\}\;,
\end{align}
\end{subequations}
with $c_{1}$ and $c_{2}$ integration constants and $\lambda_{\pm}$ given by
\begin{equation}
\lambda_{\pm} = \frac{\sqrt{1 \pm (2k_{\pm}R)^{2}}}{2}\;.	
\end{equation}
The constants $c_{1}$ and $c_{2}$ are determined by the lateral pressure exerted at the boundaries. More importantly, Eqs.~\eqref{eq:barrels_sol} reduce to simple trigonometric functions in the fine structure limit. This can be readily shown by noticing that, for $s/R \approx 1/\sqrt{K_{0}R^{2}} \ll 1$, the Laplace-Beltrami operator given in Eq.~\eqref{eq:laplace_beltrami} reduces to the second derivative with respect to $s$: i.e. $\nabla^{2}\approx \partial_{s}^{2}$. Using this in Eq.~\eqref{eq:homogeneous} and solving the resulting equation with boundary conditions $\Pi(-L/2)=\Pi(L/2)=\Pi_{\rm ext}$, with $\Pi_{\rm ext}$ the lateral pressure applied at the rims, readily yields the approximate solution
\begin{equation}
\Pi \approx \Pi_{p}+(\Pi_{\rm ext}-\Pi_{p})\,\frac{\cos(k_{\pm}s)}{\cos \left(\frac{k_{\pm}L}{2}\right)}\;.
\end{equation}

In the case of toroidal metashells, all terms in the expansion given by Eq.~(6) are non-trivial and must be computed depending on the required degree of precision. As the fine structure limit is approached, however, the series rapidly converges towards its asymptotic form $\Pi_{p}\approx-K/K_{0}$ and the particular solution can be again approximated via the $n=0$ term. That is
\begin{equation}
\Pi_{p} \approx -\frac{1}{1+\rho K_{0}\left[\rho+R\sec\left(\frac{s}{R}\right)\right]}\;.	
\end{equation}
Furthermore, when restricted to closed surfaces and in the absence of specific constraints on the global pressure, Eq.~\eqref{eq:homogeneous} always admits the trivial solution $\Pi_{h}=0$. Because the elastic energy density is proportional to $\Pi^{2}$, such a trivial solution also corresponds to the lowest energy configuration of the metashell. In spherical and pesudospherical barrels, on the other hand, $\Pi_{p}=0$ is not a solution of the boundary value problem unless $\Pi_{\rm ext}=\Pi_{p}$. 

\section{\label{sec:numerics}Numerical simulations of kirigami metashells}

Kirigami metashells are modelled as spring networks, whose topology aims at mimicking the specific combination of rigid faces and flexible joints characteristic of kirigami structures (see Fig.~\ref{fig:numerics}a). To this end, a square lattice of Hookean springs of elastic constant $k_{1}$ and rest-length $l$ is initially projected on a cylinder of radius $r$ and height $h$ (see Fig.~\ref{fig:numerics}c) and every other unit cell is rigidified by linking the vertices across the diagonals with two additional springs having rest-length  $\sqrt{2}\,l$ and the same elastic constant $k_{1}$ as the rest of the network. To implement the spatial constraint resulting from the curved geometry of the substrate, the Hooke energy of each vertex is augmented with a term proportional to the square of a function $F$, such that $F=0$ is the substrate's implicit equation. The total energy of the network is, therefore, given by
\begin{equation}\label{eq:spring_energy}
E 
= \frac{k_{1}}{2}\sum_{\langle \alpha\beta \rangle}\Bigl(|\bm{r}_{\alpha}-\bm{r}_{\beta}|-l_{\alpha\beta}\Bigr)^{2}
+ \frac{k_{2}}{2}\sum_{\alpha}F^{2}(\bm{r}_{\alpha})\;,
\end{equation}
with $k_{2}$ another elastic constant, generally larger than $k_{1}$, and $l_{\alpha\beta}$ is either $l$ or $\sqrt{2}\,l$ depending on whether the $\alpha-$th and $\beta-$th vertices are connected by the spring along an edge of a diagonal. 

The three surfaces considered in the main text consists of a spherical barrel, a pseudospherical barrel and a torus. For the former two surfaces
\begin{subequations}\label{eq:sphere_pseudo}
\begin{gather}
F_{\rm sphere} = x^{2}+y^{2}+z^{2}-R^{2}\;,\\[5pt]
F_{\rm pseudo} = z^{2}+R^{2}E^{2}\left(i\arccosh\frac{\sqrt{x^{2}+y^{2}}}{\rho}\bigg|-\frac{\rho^{2}}{R^{2}}\right)\;,
\end{gather}
\end{subequations}
To facilitate comparison between the spherical and pseudospherical shells, these have the same radius of curvature in all numerical data displayed in the main text. Analogously, for toroidal substrates one has
\begin{equation}
F_{\rm torus} = \left(R-\sqrt{x^{2}+y^{2}}\,\right)^{2}+z^{2} - \rho^{2}\;.
\end{equation}
In this case, $R$ and $\rho$ are, respectively, the distance of the torus center-line from the $z-$axis and the cross-sectional radius of the torus (see Fig.~\ref{fig:numerics}d), while the Gaussian curvature is given by
\begin{equation}
K = \frac{\cos\theta}{\rho(R+\rho\cos\theta)}\;,	
\end{equation}
where $\theta=\arcsin(z/\rho)$ is the latitude as measured from the equatorial plane, so that $\theta=0$ and $\theta=\pi$ correspond to the external and internal equators. 

\section{Floquet theory of periodic metatubes}

In the case of a periodic metatube, such as that illustrated in Fig.~3a of the main text, the force balance relation, Eq.~(5), takes the form
\begin{equation}\label{eq:metatubes}
\beta \nabla^{2}\Pi+(K+K_{0})\Pi + K = 0\;,	
\end{equation}
where the Gaussian curvature $K$ is now a periodic function of the geodesic distance $s$: i.e. $K(s)=K(s+a)$, with $a$ the geodesic length of the repeated subunits. Furthermore, within each subunit
\begin{equation}\label{eq:periodic_potential}
K = 
\left\{
\begin{array}{lll}
-1/R_{-}^{2}  & & 0 < s < b\\[5pt]
1/R_{+}^{2}  & & b < s < a\;.
\end{array}
\right.\;,
\end{equation}
with $R_{\pm}$ the curvature radii of the two regions within each subunit (i.e. yellow and green in Fig.~3). In the fine structure limit, where $K_{0}R_{\pm}^{2}\gg 1$, one can approximate $\nabla^{2}\approx\partial_{s}^{2}$ so that Eq.~\eqref{eq:metatubes} reduces to an inhomogeneous version of the classic Kroning-Penney model of electrons in a periodic potential. That is
\begin{equation}
\left(\partial_{s}^{2}+\frac{K+K_{0}}{\beta}\right)\Pi = -\frac{K}{\beta}\;.
\end{equation}
The latter, in turn, belongs to a broader class of ordinary differential equations with periodic coefficients known as inhomogeneous Hill equation and whose generic form is given in $d$ dimensions by
\begin{equation}\label{eq:hill}
\partial_{s}\bm{x} = \bm{A}\cdot\bm{x}+\bm{f}\;,	
\end{equation}
where $\bm{A}(s+a)=\bm{A}(s)$ a $d-$dimensional periodic matrix-valued function of the time like variable $s$ and $\bm{f}(s+a)=\bm{f}(s)$ a periodic source term. This class of differential equation was thoroughly investigated in Ref.~[\href{http://e-ndst.kiev.ua}{34}] using an extension of Floquet theory~[33]. An exact solution of Eq.~\eqref{eq:hill} is available in the form
\begin{equation}\label{eq:hill_solution}
\bm{x}(s) = \bm{X}(s)\cdot\left[\bm{x}(0)+\int_{0}^{s}{\rm d}s'\,\bm{X}^{-1}(s')\cdot\bm{f}(s')\right]\;.
\end{equation}
where $\bm{X}=\bm{X}(s)$ is the so-called fundamental matrix solution of the associated homogeneous equation, that is
\begin{equation}
\partial_{s}\bm{X}=\bm{A}\cdot\bm{X}\;.	
\end{equation}
Each column of the fundamental matrix solution is then an independent solution Eq.~\eqref{eq:hill} with initial conditions $\bm{X}(0)=\mathbb{1}$. Now, because of the periodic structure of Eq.~\eqref{eq:hill}, if $\bm{X}(s)$ is a solution so is the translated function $\bm{X}(s+a)$. Furthermore, these two solutions are related by the following linear transformation~[33]:  
\begin{equation}\label{eq:recursion}
\bm{X}(s+a) = \bm{X}(s)\cdot\bm{X}(a)\;.
\end{equation}
As proven in Ref.~[\href{http://e-ndst.kiev.ua}{34}], the recursive structure entailed in Eq.~\eqref{eq:recursion} implies that the asymptotic stability of the solution given in Eq.~\eqref{eq:hill_solution} depends on the spectrum of the matrix
\begin{equation}
\bm{X}(a) = e^{\bm{F}a}\;,
\end{equation}
with $\bm{F}$ a matrix. The eigenvalues of $\bm{X}(a)$ are known as Floquet multipliers and can be expressed in the form $\lambda_{n}=e^{\mu_{n}a}$ with $n=1,\,2\ldots\,d$, while $\mu_{n}$ are the so-called Floquet exponents controlling the behavior of the fundamental matrix solution $\bm{X}(s)$ in the limit $s\to\infty$.

Now, in our case, $\bm{x}=(\Pi_{h},\partial_{s}\Pi_{h})$, with $\Pi_{h}$ the solution of the homogeneous equation
\begin{equation}\label{eq:kroning_penney}
\left(\partial_{s}^{2}+\frac{K+K_{0}}{\beta}\right)\Pi_{h} = 0\;,
\end{equation}
or, equivalently
\begin{equation}
\partial_{s}
\begin{pmatrix}	
\Pi_{h}\\[5pt]
\partial_{s}\Pi_{h}
\end{pmatrix}
= 
\begin{pmatrix}	
0 						& 1\hspace{5pt}\\[5pt]
-\frac{K+K_{0}}{\beta}	& 0\hspace{5pt}
\end{pmatrix}
\cdot
\begin{pmatrix}	
\Pi_{h}\\[5pt]
\partial_{s}\Pi_{h}
\end{pmatrix}\;.
\end{equation}
The fundamental matrix solution is, therefore, of the form
\begin{equation}
\bm{X} = 
\begin{pmatrix}
\Pi_{h}^{(1)} & \Pi_{h}^{(2)} \\
\partial_{s}\Pi_{h}^{(1)} & \partial_{s}\Pi_{h}^{(2)}	
\end{pmatrix}\;,	
\end{equation}
where $\Pi_{h}^{(1)}$ and $\Pi_{h}^{(2)}$ are independent solutions of Eq.~\eqref{eq:kroning_penney} with initial conditions 
\begin{subequations}\label{eq:floquet_boundary_conditions}
\begin{gather}
\Pi_{h}^{(1)}(0) = 1\;,\qquad \partial_{s}\Pi_{h}^{(1)}(0) = 0\;,\\[5pt]
\Pi_{h}^{(2)}(0) = 0\;,\qquad \partial_{s}\Pi_{h}^{(2)}(0) = 1\;.	
\end{gather}
\end{subequations}
Eq.~\eqref{eq:kroning_penney} is formally identical to the Schr\"odinger equation of the Kronig-Penney model~[\href{http://dx.doi.org/%20https://doi.org/10.1098/rspa.1931.0019}{32}] and the solutions $\Pi_{h}^{(1)}$ and $\Pi_{h}^{(2)}$ can be readily found in the form
\begin{subequations}\label{eq:bloch_waves}
\begin{gather}
\Pi_{h}^{(1)}(s) = 
\left\{
\begin{array}{lll}
\cos (k_{-}s) 																		& \hspace{15pt} & 0 < s < b\\[5pt]
\cos (k_{-}b) \cos[k_{+}(s-b)] - \frac{k_{-}}{k_{+}}\,\sin(k_{-}b)\sin[k_{+}(s-b)]	& \hspace{15pt} & b < s < a
\end{array}
\right.\;,\\[10pt]
\Pi_{h}^{(2)}(s) = 
\left\{
\begin{array}{lll}
\frac{1}{k_{-}}\,\sin (k_{-}s) 																	& & 0 < s < b\\[5pt]
\frac{1}{k_{-}}\,\sin(k_{-}b)\cos[k_{+}(s-b)] + \frac{1}{k_{+}}\,\cos(k_{-}b)\sin[k_{+}(s-b)]	& & b < s < a
\end{array}
\right.\;,
\end{gather}
\end{subequations}
where $k_{\pm}^{2}=(K_{0}\pm 1/R_{\pm}^{2})/\beta$ and all the integration constants are chosen to guarantee the continuity of the solution and its derivative at $s=b$. From Eqs.~\eqref{eq:bloch_waves} it follows that the eigenvalues of the matrix $\bm{X}(a)$ have the form $\lambda_{\pm}=(\alpha\pm\delta)/2$, with 
\begin{equation}
\alpha=2\cos(k_{-}b)\cos[k_{+}(a-b)]-\left(\frac{k_{+}}{k_{-}}+\frac{k_{-}}{k_{+}}\right)\sin(k_{-}b)\sin[k_{+}(a-b)]\;,
\end{equation}
and $\delta=\sqrt{\alpha^{2}-4}$~[35]. In order to construct a solution of Eq.~\eqref{eq:kroning_penney} in the Bloch form, one can now assume periodic boundary conditions after an arbitrary number $N$ of repetition of the same unit cell, then Eq.~\eqref{eq:floquet_transformation} implies
\begin{equation}\label{eq:floquet_transformation}
\bm{X}(s) = \bm{X}(s)\cdot\bm{X}^{N}(a)\;.
\end{equation}
The eigenvalues $\lambda_{\pm}$ must therefore satisfy the condition $(\lambda_{\pm})^{N}=1$. In the limit $N\to\infty$, the only solutions of this equation consistent with a bounded configuration of the lateral pressure are then of the form $\lambda_{\pm}=e^{\pm ika}$, with
\begin{equation}\label{eq:floquet_exponent}
k = \frac{1}{a}\,\arctan\left(\frac{\sqrt{4\alpha^{2}-1}}{\alpha}\right)\;.
\end{equation}
Together with Eq.~\eqref{eq:hill_solution}, this implies that the general solution $\Pi_{h}$ of the homogeneous equation associated to Eq.~\eqref{eq:metatubes} satisfies $\Pi_{h}(s+a)=e^{ika}\Pi_{h}(s)$, thus
\begin{equation}\label{eq:bloch}
\Pi_{h}(s) = e^{iks}u_{k}(s)\;	
\end{equation}
with $u_{k}$ a periodic function of period $a$: i.e. $u_{k}(s)=u_{k}(s+a)$. The band structure shown in Fig.~3c, in particular, is obtained by first calculating the eigenvalues of $\bm{X}(a)$ for a specific $K_{0}$ value and then computing $k$ from Eq.~\eqref{eq:floquet_exponent}.

%% file: main.bbl
%